\documentclass[preprint,11pt,3p]{elsarticle}
\makeatletter
\def\ps@pprintTitle{%
  \let\@oddhead\@empty
  \let\@evenhead\@empty
  \let\@oddfoot\@empty
  \let\@evenfoot\@oddfoot
}
\usepackage{color}
\definecolor{light-gray}{gray}{0.80}
\usepackage[]{listings}
\usepackage{xcolor}
\usepackage{amsfonts}
\usepackage{amsmath}
\usepackage{amssymb}
\usepackage{bm}
\usepackage{textcomp,mathcomp}
\usepackage{pgfgantt}
\usepackage{tikz}
\usepackage{caption}
\usepackage{subcaption}
\usepackage[colorlinks,citecolor=red,urlcolor=blue,bookmarks=false,hypertexnames=true]{hyperref} 
\usepackage[linesnumbered,ruled,vlined]{algorithm2e}
\usepackage{algorithmic}
\usepackage{multirow}
\usepackage{graphicx}
\usepackage{lineno}
\usepackage{enumerate}
\usepackage[mathscr]{euscript}
\usepackage{multirow}
\usepackage{epstopdf}
\usepackage{tikz}
\tikzstyle{startstop} = [rectangle, rounded corners, minimum width=3cm, minimum height=1cm, align=center, text width=4cm, draw=black, fill=red!30 ]
\tikzstyle{io} = [trapezium, trapezium left angle=70, trapezium right angle=110, minimum width=3cm, minimum height=1cm, align=center, text width=4cm, draw=black, fill=blue!30]
\tikzstyle{process} = [rectangle, minimum width=3cm, minimum height=1cm, align=center, text width=4cm, draw=black, fill=orange!30]
\tikzstyle{process1} = [rectangle, minimum width=5cm, minimum height=1cm, align=center, text width=5cm, draw=black, fill=orange!30]
\tikzstyle{decision} = [diamond, minimum width=1cm, minimum height=1cm, align=center, text width=4cm, draw=black, fill=green!30]
\tikzstyle{arrow} = [thick,->,>=stealth]

\graphicspath{{}}
\usepackage{xcolor}
\usepackage{soul}
\usepackage{subfiles}
\renewcommand\hl[1]{\ignorespaces}

\usepackage[export]{adjustbox}
\renewcommand{\hl}[1]{#1}

\begin{document}

\begin{frontmatter}

\title{FeynKrack: A continuum model for quasi-brittle damage through Feynman-Kac killed diffusion}



\hypersetup{pdfauthor={Name}}
\author[1]{Ved Prakash}
\author[1]
{Upadhyayula M. M. A. Sai Gopal}
\author[2]
{Sanhita Das}
\author[1]
{Ananth Ramaswamy}
\author[1]{Debasish Roy\corref{cor1}}\ead{royd@iisc.ac.in}
\address[1]{Department of Civil Engineering; Indian Institute of Science,Bangalore 560012, India}
\address[2]{Department of Civil and Infrastructure Engineering; Indian Institute of Technology Jodhpur, Rajasthan 342030, India}
\cortext[cor1]{Corresponding author}
\begin{abstract}
{Continuum damage mechanics (CDM) is a popular framework for modelling crack propagation in solids. The CDM uses a damage parameter to quantitatively assess what one loosely calls `material degradation'. While this parameter is sometimes given a physical meaning, the mathematical equations for its evolution are generally not consistent with such physical interpretations. Curiously, degradation in the CDM may be viewed as a change of measures, wherein the damage variable appears as the Radon-Nikodym derivative. We adopt this point of view and use a probabilistic measure-valued description for the random microcracks underlying quasi-brittle damage. We show that the evolution of the underlying density may be described via killed diffusion as in the Feynman-Kac theory. Damage growth is then interpreted as the reduction in this measure over a region, which in turn quantifies the disruption of bonds through a loss of force-transmitting mechanisms between nearby material points. Remarkably, the evolution of damage admits an approximate closed-form solution. This brings forth substantive computational ease, facilitating fast yet accurate simulations of large dimensional problems. By selecting an appropriate killing rate, one accounts for the irreversibility of damage and thus eliminates the need for ad-hoc history-dependent routes typically employed, say, in phase field modelling of damage. Our proposal FeynKrack (a short form for Feynman-Kac crack propagator) is validated and demonstrated for its efficacy through several simulations on quasi-brittle damage. It also offers a promising stochastic route for future explorations of non-equilibrium thermodynamic aspects of damage.}
\end{abstract}
\end{frontmatter}


\section{Introduction}
Quasi-brittle failure in materials such as concrete, rocks, ceramics, etc. is typically preceded by an intricate web of random microcracks. A faithful prediction of such complex crack patterns at the mesoscale is generally infeasible and one is thus more interested in models that can reproduce the propagation of quasi-brittle damage at the macroscale. However, except for a few attempts \cite{thamburaja2021fracture}, most available models for quasi-brittle damage do not have the physical motivation or mathematical grounding necessary for high predictive fidelity to replace repetitive and costly experiments. Given the inherent randomness in the distribution of microcracks, a model that exploits a probabilistic measure-valued description of microcracks seems to be a possibility. In this context, we note that the degradation function used with continuum damage mechanics (CDM) \cite{lemaitre1978aspect} may be looked upon as a Radon Nikodym derivative associated with a change of probability measures. We plan to follow this thread further in this article.  

Initially, linear elastic fracture mechanics (LEFM) and elasto-plastic fracture mechanics (EPFM) formed the basis for many continuum models on fracture and damage. But in the absence of proper length scales, they could not model the geometric and material non-linearities in the process-zone mechanics, an aspect considered crucial to quasi-brittle fracture \cite{bazant2019fracture}. Moreover, fracture preceded by degradation processes not necessarily arising due to mechanical loads cannot be captured by LEFM or EPFM. In contrast, damage mechanics \cite{lemaitre1978aspect} equipped with proper variables and length scales not only modelled the damage phenomena, but also the continuous material degradation as a precursor to crack propagation. 

Damage models may be discrete or continuous. While a discrete approach allows for a jump in the deformation field due to the presence of a crack, the jump is smeared out in a continuous approach. Discrete approaches require prior knowledge of the crack path and thus pose challenges in numerical implementation. Several enhancements in discrete approaches have been proposed, e.g. FEM with remeshing, XFEM, and cohesive- zone \cite{moes1999finite,belytschko1999elastic} requiring local enrichment and complex crack-tracking algorithms. But they still remain highly mesh-dependent and computationally intensive. Further developments such as discontinuous FEM \cite{mergheim2006geometrically}, inter-element separation model \cite{xu1996numerical}, cracking particle method  \cite{rabczuk2004cracking}, though significant, fall short of expectations in modelling complex crack trajectories, especially in three-dimensional fracture.

Continuum damage models are superior to discrete approaches as they do not require a prior knowledge of the crack path and are numerically more stable. Most of these models are phenomenological. For example, the size effect model and the two-parameter fracture model \cite{morgan1997detecting} rely on crack-width, stress intensity factors with no reference to underlying bond-breakage or stiffness degradation. Microplane and crack band models \cite{bavzant1983crack,bavzant1988microplane} are two approaches widely used in modelling quasi-brittle fracture. Though they accurately capture damage orientation, damage evolution laws often violate the laws of thermodynamics. In modelling damage orientation, however, they are superior to second and fourth-order tensor-based continuum approaches as constitutive equations are defined using vectors representing stress and strain on randomly oriented planes.  

Of most local continuum models, approaches based on continuum damage mechanics (CDM) invoke the concept of a damage variable describing the physics of underlying degradation. These models are thus physically more meaningful. Damage variables with different interpretations have been used, such as measuring  effective load-carrying areas \cite{kachanov2013introduction}, variations in elastic modulus \cite{lemaitre1978aspect}, void volume fractions in ductile materials \cite{gurson1977continuum}, microcrack density \cite{ulloa2022micromechanics}, post-rupture bond-density, nucleation-growth-coalescence of microvoids etc. We find scalar, vector, or higher-order tensor representations of these damage variables \cite{murakami1987progress} carrying information on degradation and its orientation. The key to successfully capture complex crack paths with the CDM is by constructing evolution equations consistent with the adopted interpretation of the damage variable. This important feature is however conspicuous by its absence in most models. 

CDM models also lack numerical stability unless length scales are incorporated. Using length scales, non-local and hence regularized damage models with length scales have been prescribed \cite{peerlings1998gradient} so as to prevent localisation of damage to zero-volume, mesh dependence and unphysical softening  \cite{pijaudier1987nonlocal, bavzant1989measurement}. Even from a physical standpoint, non-locality makes sense as damage at a material point might indeed depend on its neighbourhood \cite{bavzant1991continuum}, especially via interactions among microcracks at the mesoscale.

Both gradient and integral-based non-local models have been used for quasi-brittle fracture; but they have their drawbacks. Gradient-based models essentially originate from integral approaches \cite{peerlings1998gradient}. Among these, phase-field damage models are widely used for quasi-brittle fracture \cite{wu2017unified}. Despite several advantages, it suffers from crucial drawbacks such as opaqueness in a physical interpretation of phase field, unphysical energy functionals, high computational cost, mesh-sensitivity \cite{zhang2017numerical} etc. The governing equation is ill-conditioned for soft materials, dynamic fracture, multi-modal fracture and material and geometric nonlinearities. To address these challenges, elaborate measures are typically required \cite{kopanivcakova2023nonlinear}, possibly at an additional computational cost. In some cases, numerical issues arise from the monolithic solving of the balance of linear momentum and damage evolution. To prevent this, an arbitrary history variable is introduced \cite{miehe2010phase}. 

The high sensitivity of gradient-type models to algorithmic parameters is, to an extent, eliminated in integral-type models. Peridynamics \cite{wu2020rate} is a widely used integral model, but with questionable accuracy \cite{gu2019possible}. Efforts to enhance it have increased its complexity \cite{madenci2019weak}. \hl{In the context of quasi-brittle fracture, various attempts to improve the peridynamics have been made recently (e.g.  \mbox{\cite{friedrich2025implementation})}, and the references therein}. Other non-local prescriptions include \cite{lv2024energy} claiming better macro-meso-scale consistency. \hl{Lattice discrete element method is also gaining popularity in modelling quasi brittle fracture \mbox{\cite{kosteski2020size}}}. Recently, a nonlocal damage model within the graph-based finite element analysis (GraFEA) has been proposed for concrete \cite{thamburaja2019fracture}, drawing inspiration from the microplane theory in \cite{bazant2019fracture}. Crack closure and rate dependence are key features of this model. Evolution of the damage variable, which represents the survival probability of the crack plane, is governed by a master equation. This equation however does not follow any of the known probability evolution, such as Liouville, or Fokker-Planck or Feynman-Kac. 

The current work relates damage to a continuous measure, such that over any region in the material body, it tracks the remaining or intact 'macroscopic bonds' through which forces could be transferred between material points. Accordingly, the field variable that relates to damage in this model is this measure over any subregion of unit volume, i.e. it is also the probability density (or the Radon-Nikodym derivative with respect to Lebesgue volume measure) for the macroscopic bonds that are intact. Despite exploiting aspects of probability theory, specifically the Feynman-Kac path integral for killed diffusion \cite{klebaner2012introduction,holcman2005survival,premkumar2023generative}, this model does not need any pathwise simulations involving the explicit use of noise. In other words, the governing equations continue to be PDEs and, in that sense, our approach might as well be considered as a variant of the non-local CDM. However, unlike most other models, the evolution of the probability characterizing damage is now consistent with our physical interpretation of damage. Additionally, this evolution has a closed-form expression, which greatly reduces the computational effort and allows for a much higher time-stepping in evolving damage. We use this model, codenamed FeynKrack, to numerically investigate several problems of crack propagation, the first being that in a concrete specimen in a three-point bending test. For mixed-mode fracture in 2D, we simulate  L-shaped plate and a square plate with a tilted crack. We predict wing cracks in a 3D solid with through-crack. Moreover, crack propagation in 3D Brazilian test specimens with an inclined notch is also predicted successfully.  

The rest of this article is structured as follows. In \S \ref{sec:Motivation}, we discuss the underlying rationale for our approach to reflect on the connection between degradation and a probability measure. We also briefly overview killed diffusion processes and explain our choice of this specific stochastic process. The derivation of the governing PDE and dissipation inequality is in \S \ref{sec:formulation}. \S \ref{sec:SpecializatioofConstitutiveFunction} outlines the constitutive equations used in our numerical implementation. In \S \ref{sec:Validation}, we first numerically study the role of model parameters and then reproduce a few key experimental observations in concrete and rock fracture. Finally, we conclude the article in \S \ref{sec:conclusion}.

\section{FeynKrack: Motivation and foundational concepts}
\label{sec:Motivation}
\subsection{A probabilistic makeover of damage}\label{sec:MotivationRadon}
We assume that the deformation of the material body can be deterministically described in the intact (i.e. undamaged or pristine) state and that damage initiates by the formation of random microcracks leading to progressive snapping of 'macroscopic bonds' responsible for mechanical force transfer among the material points. We use a continuous probability measure $P$, defined as part of a complete probability space $(\Omega, \mathscr{F}, P)$, to describe these bonds and their evolution during damage. Here $\Omega$ denotes the sample space consisting of possible number densities of these bonds and $\mathscr{F}$ the associated Borel $\sigma  $-algebra consisting of (all unions and complements of) open subsets of $\Omega$ including the null set $\phi$ and $\Omega$ itself.

CDM postulates that the free energy density $\psi_d$ of a damaged solid is related to $\psi_0$, the free energy of the intact body through a degradation function $w$, i.e. $\psi_d = w \psi_0$. Integrated over the material domain $\Lambda_0$, the free energy in the damaged body is given by 
$\int_{\Lambda_0} \psi_d d\Lambda_0$, $d \Lambda_0$ being the Lebesgue incremental volume measure (a 3-form). Let $P(d{\Lambda}_0)$ be the modified incremental measure due to damage over the same domain. Since the total energy in the damaged material is invariant with respect to an absolutely continuous change of measures, the following must hold true,
\begin{equation}
    E_d = \int_{\Lambda_0} \psi_d d\Lambda_0 = \int_{\Lambda_0} \psi_0 P(d\Lambda_0) = \int_{\Lambda_0} \psi_0 w d\Lambda_0
    \label{eq:ModElasEnergy}
\end{equation}
where 
\begin{equation}\label{RadonNikodym}
w=\frac{P(d\Lambda_0)}{d\Lambda_0}=\frac{dP(\Lambda_0)}{d\Lambda_0}
\end{equation}
$w$ may be interpreted as a Radon-Nikodym derivative. 
We introduce a probability density $g$ over $\Omega$, corresponding to the density of $P$ and describing the number density (per unit volume) of macroscopic bonds. In the absence of damage, the measure $P$ is the trivial identity, i.e. $P(d\Lambda_0)=d\Lambda_0$. At this stage, microcracks have not yet formed at a material point $\boldsymbol{x}$ so that the issue of non-local interaction with a degraded neighborhood does not arise. However, as microcracks form at $\boldsymbol{x}$ and spread as a function of time, the associated density $g$ has a non-uniform (e.g. hump-like) structure around $\boldsymbol{x}$ whilst continuing to decrease, perhaps until a critical limit is reached when the body is macroscopically (i.e. visibly) fractured. Accordingly, $g$ should not be treated as a normalizable probability density so that the progressive decrease in the macroscopic bond density could be modeled through an appropriate killing rate. However, in the absence of any prior constraints based on the mechanics of deformation and damage, the \textit{a-priori} evolution of microcracks and hence the loss of macroscopic bonds is assumed as purely diffusive. In such a case, the density $g$ satisfies the diffusion equation, i.e. the Fokker-Planck equation for 3D Brownian motion. This is clearly unphysical, as microcracks do not usually diffuse through the solid like a drop of ink does in water. Our idea is to introduce an extra term into this evolution, which signifies a killing rate of the measure of macroscopic bonds consistent with the known mechanics of quasi-brittle damage. 

\subsection{Nonlocal degradation}\label{sec:MotivationNonlocaloperator}

The presence of a continuous measure makes the deformation accompanied by damage necessarily nonlocal. To determine the nonlocal forms for the stresses and strains, we define the integral operator $\mathscr{R}$ over a subset $\mathrm{B}\subset\Lambda_0$ as,
\begin{equation}\label{AveragingOp}
    \mathscr{R}(A(\boldsymbol{x})) =\int_{\mathrm{B}} A(\boldsymbol{x}^{\prime}) dP(\mathrm{B}) =\int_{\mathrm{B}} A(\boldsymbol{x}')g\left(\lvert\boldsymbol{x}'-\boldsymbol{x}\rvert\right)   d\boldsymbol{x}^{\prime} 
\end{equation}
$\mathscr{R}$ determines the average or mean or the first moment (under the measure $P$) of the field $A$ evaluated at $\boldsymbol{x}$. The size of the subset $\mathrm{B}$ is related to the support of $g$ and, for pure diffusion, it is proportional to $\sqrt{\Delta t}$, the time over which micro-cracks have evolved. In writing the non-local operator as above, we have assumed symmetry in the density of $P$, i.e. 
\begin{equation}
g\left(\boldsymbol{x}'-\boldsymbol{x}\right) = g\left(\boldsymbol{x}-\boldsymbol{x}'\right) = g\left(\lvert\boldsymbol{x}'-\boldsymbol{x}\rvert\right)
\end{equation}
Indeed, this form of the density $g$ of $P$ resembles a delocalisation kernel typically employed in nonlocal elasticity \cite{polizzotto2001nonlocal}. However, conceptually and functionally, it fundamentally differs from a conventional delocalisation kernel. Moreover, it is different from the various evolution equations attempted in the past \mbox{\cite{bai2005statistical}}. As will be shortly shown, it is related to the transition probability density of a killed diffusion process with a state-dependent killing rate derived based on the known physics of quasi-brittle fracture. The so called radius of nonlocal interactions in our case is decided by the intensity of the Brownian motion in 3D used in defining the prior measure.   

To motivate the detailed expression for $g$, the linkage between bond-breaking and killing (i.e. a gradual decrease) of Brownian measure for bonds must be understood. Since we are dealing with a continuous measure, consider a realized set of macroscopic bonds in a region (around the point $\boldsymbol{x}$) containing straight lines connecting $\boldsymbol{x}$ and $\boldsymbol{x}'$ (for different $\boldsymbol{x}'$). The measure of bonds over this region reduces from its value for the intact material if there are intervening microcracks affecting the transfer of mechanical forces. The killing rate, which describes this reduction, applies the necessary control on an otherwise Brownian scatter of the microcracks or the loss of macroscopic bonds. We elaborate on an expression of this killed measure in the following section.

\subsection{Killed diffusion process}\label{sec:MotivationKilledDiffusion}
We have been emphasizing on the Brownian measure imposed a-priori on the decreasing macroscopic bonds. Thus, if we evolve a so-called \lq{particle\rq}  executing Brownian motion, it will chart out a path (trajectory) on which any two material points would suffer a loss of macroscopic bonding. Such Brownian dispersion of snapped bonds cannot continue indefinitely without allowing damage to spread all over the entire solid body. Thus, we need to kill or terminate this Brownian trajectory with a certain probability.

Consider the macroscopic bonds between the point $\boldsymbol{x}$ and the infinitesimal region $d \boldsymbol{x}'$. As before, we keep $\boldsymbol{x}$ fixed and let the neighbors $\boldsymbol{x}'$ interact with $\boldsymbol{x}$ by transmission of mechanical force. Degradation and eventual snapping of these bonds, leading to a loss in the transmitted force, is a stochastic process. Consider a vector valued stochastic process $\boldsymbol{\alpha}(\boldsymbol{x},\tau)\in \mathbb{R}^3$ suitably defining the loss in the maximal traction that can be transmitted across unit areas of the three orthonormal planes passing through the point $\boldsymbol{x}$ at time $\tau$. In general, we may write a stochastic differential equation (SDE) for $\boldsymbol{\alpha}$ as $d\boldsymbol{\alpha}(\boldsymbol{x},\tau) = \boldsymbol{F}(\boldsymbol{x},\tau) d\tau + \beta(\boldsymbol{x},\tau) d\boldsymbol{B}$. The forcing term $\boldsymbol{F}$ denotes body forces arising from gravitational or electromagnetic effects and $\boldsymbol{B}_\tau\in\mathbb{R}^3$ is a vector-valued standard Brownian motion, say, with a scalar noise intensity $\beta(\boldsymbol{x},\tau)$. The noise term corresponds to a Brownian scatter of the vector-valued degradation process $\boldsymbol{\alpha}$. For a given point $\boldsymbol{x}$ and its neighbour $\boldsymbol{x'}$, the transition probability $P(\boldsymbol{\alpha}(d\boldsymbol{x'}),t|\boldsymbol{\alpha}(\boldsymbol{x}),\tau)$, $\tau\le t$, for evolving $\boldsymbol{\alpha}$ is the incremental measure for degradation and the associated density $p(\boldsymbol{\alpha}(\boldsymbol{x'}),t|\boldsymbol{\alpha}(\boldsymbol{x}),\tau)$ satisfies the backward Kolmogorov equation in $\boldsymbol{x}, \tau$. Equivalently, it also satisfies the adjoint form of this backward equation, called the Fokker-Planck equation, in $\boldsymbol{x'}, t$. To see this, one may consider an arbitrary scalar-valued (and Borel measurable) function $H(\boldsymbol{\alpha},\tau)$ and use Ito's formula along with the SDE for $\boldsymbol{\alpha}$ to write an SDE for $H$. By taking expectations of the terms in this SDE, we may get a PDE for $\mathbb{E}(H)$, where $\mathbb{E}$ denotes the expectation operator. Taking  $H(\boldsymbol{\alpha},\tau)$ to be the indicator function over a subset $A\subset \Omega$, taking the expectation, dividing by the Lebesgue measure $\textrm{meas}(A)$ and taking limit as $\textrm{meas}(A)\rightarrow 0$, we may arrive at the required backward equation for the probability density. The Fokker-Planck equation follows by taking the adjoint of the backward equation. An alternative way to derive the Fokker-Planck equation is through Kramer-Moyal expansion \mbox{\cite{luczka1995non}}.

However, when damage is isotropic as we presently assume, the three scalar components of $\boldsymbol{\alpha}$ are the same, i.e. $\alpha_1=\alpha_2=\alpha_3=a(\boldsymbol{x},\tau)$. In this case, the three components of the vector field $\boldsymbol{F}$ are also the same ($\boldsymbol{F}=\boldsymbol{0}$ in this work). Given an oriented plane through $\boldsymbol{x}(\tau)$ and the associated $a$, let $\hat\nu$ be the unit vector along which $a$ acts. Taking a neighboring point $\boldsymbol{x'}(t)$, such that the so-called bond $\boldsymbol{x'}-\boldsymbol{x}$ is along $\hat\nu$, the transition probability density of $a$ quantifying degradation is the same as that of the bond itself. 
To describe degradation, we may thus as well express the incremental transition measure of the bond as $P(d\boldsymbol{x}',t\lvert \boldsymbol{x},\tau)=p(\boldsymbol{x}',t\lvert \boldsymbol{x},\tau)d\boldsymbol{x}'$. The Fokker-Planck equation is: 

\begin{equation}\label{ForwardKolmogorv}
     \frac{\partial p(\boldsymbol{x}',t\lvert \boldsymbol{x},\tau)}{\partial t} =  L^{*}_tp = -\nabla \cdot (F(\boldsymbol{x}',t)p(\boldsymbol{x}',t\lvert \boldsymbol{x},\tau))+\frac{\beta^2 (t)}{2}\nabla \cdot \nabla p(\boldsymbol{x}',t\lvert \boldsymbol{x},\tau)
%
\end{equation}
where $F(\boldsymbol{x}',t)$ and $\beta^2(t)\mathcal{I}$ are respectively the drift vector and diagonal diffusion matrix, and $\mathcal{I}$ denotes the identity matrix. $\beta^2(t)$ is identifiable with the ambient temperature in a region around the point $\boldsymbol{x}$ undergoing degradation due to the evolving microcracks. $L_t\equiv F(\boldsymbol{x}',t) \cdot \nabla + \frac{\beta^2}{2}\nabla \cdot \nabla$ is the infinitesimal generator of the SDE for $\alpha$ and
$L^{*}_t$ (defined as $\int L_t (u) v d\boldsymbol{x} = \int u L^{*}_t (v) d\boldsymbol{x}$, where $u$ and $v$ are arbitrary integrable functions) is its adjoint. The corresponding backward Kolmogorov equation has the following form:

\begin{equation}\label{BackwardKolmogorov}
 \frac{\partial p(\boldsymbol{x'},t\lvert \boldsymbol{x},\tau)}{\partial \tau} = -L_\tau p = -F(\boldsymbol{x},\tau) \cdot \nabla p(\boldsymbol{x'},t\lvert \boldsymbol{x},\tau) - \frac{\beta(\tau)^2}{2} \nabla \cdot \nabla p(\boldsymbol{x'},t\lvert \boldsymbol{x},\tau)   \end{equation}

Leaving out the body forces, we are yet to account for the effect of the externally applied loading on degradation. Toward this, a generalized form of the backward Kolmogorov equation the Feynman-Kac equation, is needed, wherein an additional killing term is introduced. This term is a control on the evolution of $\boldsymbol{\alpha}$ due to applied loading and boundary conditions. It is not a part of the SDE for $\boldsymbol{\alpha}$ and the control is exercised by eliminating (i.e. killing or stopping) a fraction of the ensemble of trajectories of $\boldsymbol{\alpha}$. A carefully constructed killing  imparts a direction to the Brownian scatter consistent with the applied load. Denoting $q(\boldsymbol{x},\tau)=p(\boldsymbol{x'},t|\boldsymbol{x},\tau)$ for given $\boldsymbol{x'}$ and $t\ge\tau$, we have the following Feynman-Kac equation:

\begin{equation}\label{BackwardKilling}
     \frac{\partial q(\boldsymbol{x},\tau)}{\partial \tau} = -L_\tau q = -F(\boldsymbol{x},\tau) \cdot \nabla q(\boldsymbol{x},\tau) - \frac{\beta(\tau)^2}{2} \nabla \cdot \nabla q(\boldsymbol{x},\tau
    ) + G(\boldsymbol{x},\tau) q(\boldsymbol{x},\tau)
\end{equation}
$G(\boldsymbol{x},\tau)$ is the growth or killing rate, which controls either the growth $(G > 0)$ or termination $(G < 0)$ of bonds. When $G=0$, we set $\beta=0$ so that $q$ does not evolve (with $\boldsymbol{F}=\boldsymbol{0}$).
Thanks to the killing term, the generator $L_\tau$ is non-conservative, i.e. $L_\tau 1(\boldsymbol{x}) \neq 0$ where $1(\boldsymbol{x})$ is the function that evaluates to 1 for any $\boldsymbol{x}\in \Omega$. The corresponding forward equation may be obtained via the adjoint of $L_\tau$. Since the solution to the last equation must be interpreted as evolving backward in time, we may consider the following boundary value problem with a terminal condition, 
\begin{subequations}
    \begin{align}
     \frac{\partial q(\boldsymbol{x},\tau)}{\partial \tau} = -F(\boldsymbol{x},\tau
    ) \cdot \nabla q(\boldsymbol{x},\tau
    ) - \frac{\beta(\tau)^2}{2} \nabla \cdot \nabla q(\boldsymbol{x},\tau)+ G(\boldsymbol{x},\tau
    ) q(\boldsymbol{x},\tau)\label{eq:BackKolmKilDiff}\\
  \text{with }  q(\boldsymbol{x},t) = q_t(\boldsymbol{x}), \tau<t \label{terminalcondition}
    \end{align}
\end{subequations}

$q(\boldsymbol{x},\tau)$ has a closed-form solution given by the Feynman-Kac formula (derivation provided in supplementary material)
\begin{equation}
q(\boldsymbol{x},\tau
    ) = \mathbb{E}_P \left[ q_t(\boldsymbol{\xi})\text{exp}\left(-\int_\tau^t G (\boldsymbol{\xi}(s),s) ds)\right)  \lvert \boldsymbol{\xi}(\tau) = \boldsymbol{x} \right]
    \label{eq:FeynKacKildiff}
\end{equation}

Eq. \eqref{eq:FeynKacKildiff} may also be written as
\begin{equation}
q(\boldsymbol{x},\tau
    ) = \int_{-\infty}^{\infty}q_t(\boldsymbol{x'}) K(\boldsymbol{x'},t\lvert \boldsymbol{x},\tau) d\boldsymbol{x'} 
\end{equation}
where the kernel $K$ is defined as
\begin{equation}
K(\boldsymbol{x'},t\lvert \boldsymbol{x},\tau) = p(\boldsymbol{x'},t\lvert \boldsymbol{x},\tau) \text{exp}\left(-\int_\tau^t G (\boldsymbol{x'}(s),s) ds)\right) 
\end{equation}
If we consider the transition probability over a small time interval $\Delta t=t-\tau$, $\boldsymbol{x}'(t)$ given $\boldsymbol{x}(\tau)$ is approximately distributed as Gaussian times an exponential factor whose argument depends on the killing rate. More specifically, we have:
\begin{equation}
    K (\boldsymbol{x'}\lvert \boldsymbol{x}) = \frac{1}{\sqrt{2\pi\beta^2(t)\Delta t}}\text{exp}\left[-\frac{(\boldsymbol{x'} - \boldsymbol{x} - F(\boldsymbol{x},t)\Delta t)^2}{2\beta^2(t)\Delta t}-G(\boldsymbol{x},t)\Delta t\right]
\end{equation}
where, for notational convenience, we have used $K(\boldsymbol{x}'|\boldsymbol{x}):=K(\boldsymbol{x}', \Delta t|\boldsymbol{x})$. $K(\boldsymbol{x}'|\boldsymbol{x})$ may be looked upon as the random reduction over $\Delta t$ in the number density of macroscopic bonds when a microcrack connects $\boldsymbol{x}'$ with $\boldsymbol{x}$. If $F=0$, the number density of bonds reduces only due to the killing term (even though random fluctuations of low probability leading to an increase in the bond density is also possible). We continue to take $F=0$ in the rest of the article so that the transition density (over $\Delta t$) of the macroscopic bonds takes the form:

\begin{equation}
    g (|\boldsymbol{x}'- \boldsymbol{x}|) = \frac{1}{(\sqrt{2\pi\beta^2(t)\Delta t})^3}\text{exp}\left[-\frac{|\boldsymbol{x}' - \boldsymbol{x}|^2}{2\beta^2(t)\Delta t}-G(\boldsymbol{x},t)\Delta t\right]
\end{equation}
Indeed, when $G=0$, the above density is Gaussian with mean $\boldsymbol{x}$ and covariance matrix $\beta^2 \Delta t\boldsymbol{I}$. $\boldsymbol{I}$ denotes the $3\times 3$ identity matrix for 3D solids. 


\subsection{Further discussion on damage as a measure }\label{sec:MotivationDamageProbability}

\begin{figure}[ht!]
    \centering
\includegraphics[width=0.99\textwidth]{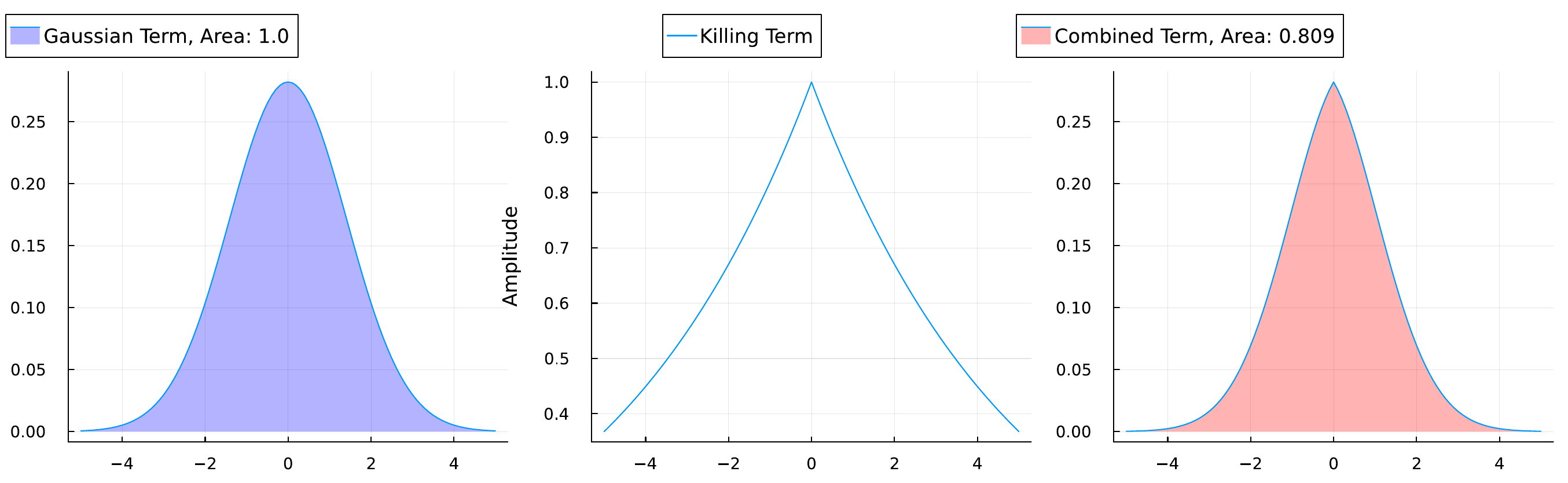}
    \caption{Characteristics of the transition density: (leftmost) plot without any killing term;  (middle) plot of an arbitrary killing term; (rightmost) plot with both terms combined, showing a reduction in the area under the curve due to killing. For the 1D calculation, parameters were set as $x = 0$, $\Delta t = 1.0$, $\beta = \sqrt{2}$; $y$-axis gives density and $x$-axis represents $x'$.}
    \label{fig:TranDenChar}
\end{figure}

Now that we have furnished killed diffusion as a descriptor of degradation or the bond-breaking process, given the initial time $t_0$, a material point $\boldsymbol{x}$ and a state-dependent killing rate $G$, the transition density $g$ of the bonds over an interval $t-t_0$ takes the form,
\begin{equation}
g\left(\lvert\boldsymbol{x}'-\boldsymbol{x}\rvert\right) = \frac{1}{(2 \pi \beta^2 (t - t_0))^{\frac{n}{2}}}\, \text{exp}\bigg\{-\int_{t_0}^{t}Gds - \frac{\lvert \boldsymbol{x}' - \boldsymbol{x}\rvert^2}{2 \beta^2 (t - t_0)}\bigg\} 
\label{eq:phitransition}
\end{equation}
In the above equation, $n=2$ or $n=3$ represents the dimension of the solid body (2D or 3D). The covariance matrix is given by $\beta^2 (t-t_0) \boldsymbol{I}$, where $\boldsymbol{I}$ is the $n\times n$ identity matrix. 
For $G = 0$, Eq. \eqref{eq:phitransition} simplifies to a Gaussian distribution characteristic of a zero-mean Brownian motion of intensity $\beta (t)$. When integrating over an infinite domain for a fixed $\boldsymbol{x}$, the integral becomes 1 for $G = 0$. For $G = 0$ and $\beta =0$, all bonds connected to $\boldsymbol{x}$ are intact representing a completely undamaged scenario. In such a case, $g$ approaches a Dirac delta at $\boldsymbol{x}$ so that its integral over any sub-region of the body containing $\boldsymbol{x}$ evaluates to 1. For $G,\textrm{ }\beta > 0$, $g$ is non-Gaussian and its integral over a sub-region gives the change (decrease) in the measure of bonds over the sub-region (see Fig. \mbox{\ref{fig:TranDenChar})}. As $G$ increases monotonically, the integral centered at $\boldsymbol{x}$ decreases monotonically toward 0, an idealistic terminal scenario where no bonds pass through $\boldsymbol{x}$. Borrowing from the classic CDM terminology and starting with the 'intact' state at time $t_0=0$, $g$ is the degradation at time $t$ and the damage variable is the integral of $g$ over the domain at that time.

A closed-form expression for $g$ yields a significant benefit in numerical simulations. Typically in a regularized continuum damage model, degradation evolves according to a partial differential equation or an integro-differential equation, which is often stiff and demands considerable computational effort for a reasonable solution. Elimination of this effort should accelerate the numerical work manifold. 

As derived in \cite{chetrite2015nonequilibrium}, it is possible to arrive at a conservative generator $\mathbb{L}$ corresponding to the generator $L$ in Eq. \eqref{BackwardKilling}, which is non-conservative owing to the killing term. This correspondence or equivalence is in the asymptotic sense as defined in the last cited reference. To derive $L$, one must solve the eigenvalue problem:
\begin{equation}\label{Eigen}
L \mathsf{r} = \lambda \mathsf{r}
\end{equation}
where $\lambda$ is an eigenvalue, say the largest or dominant one, and $r$ the corresponding right (and normalized) eigenfunction. Note that $\lambda$ is a function of the killing rate $G$. Then we may treat $\mathsf{r}$ as a Doob's h-function and write $\mathbb{L}$ via the following generalized h-transform:
\begin{equation}\label{htransform}
 L\Psi=\frac{1}{\mathsf{r}} L(\mathsf{r}\Psi)-\frac{1}{\mathsf{r}} (L\mathsf{r})\Psi 
\end{equation}
for any function $\Psi$. Starting from $\boldsymbol{x}$ at the initial time $t_0=0$ and considering the interval $(0,t)$, the path measure of the h-transformed process is approximately related to the original process as:
\begin{equation}\label{rnpath}
    \frac{dP_{\mathbb{L},t}}{dP_{L,t}}=\mathsf{r}^{-1}(\boldsymbol{x})\exp{(-\lambda t)}\mathsf{r}(\boldsymbol{x}':=\boldsymbol{x}(t))
\end{equation}
The probabilistic worldview based on the generator $\mathbb{L}$ does not have a killing term, but a drift term given by $-\beta^2 \nabla \ln r(\boldsymbol{x}')$ which has a similar (asymptotic) control on the propagating cracks as the killing term. However, what is remarkable is that the h-transformed equation now corresponds to a gradient flow with the h-function (i.e. the eigenfunction $r$)  acting as the potential (e.g. dissipation) that is being extremized (e.g. maximized) by the flow, i.e. the propagating microcracks. Clearly, $h$ is related to the so-called fracture energy release rate. In addition, the dominant eigenvalue $\lambda(G)$ is the  scaled cumulant generating function, whose Legendre-Fenchel transform gives the large deviation rate function (or dynamical entropy) for the dynamics of quasi-brittle damage. A given value of this rate function should thus reflect different microcrack configurations leading to the same macroscopic bond density. Use of this powerful probabilistic framework to develop fluctuation relations might lead to a new perspective on the non-equilibrium thermodynamics of quasi-brittle fracture; however we do not pursue it here. 

\section{Formulation}
\label{sec:formulation}
\subsection{Kinematics}\label{Kinematics}
\begin{figure}[ht!]
    \centering
\includegraphics[width=0.35\textwidth]{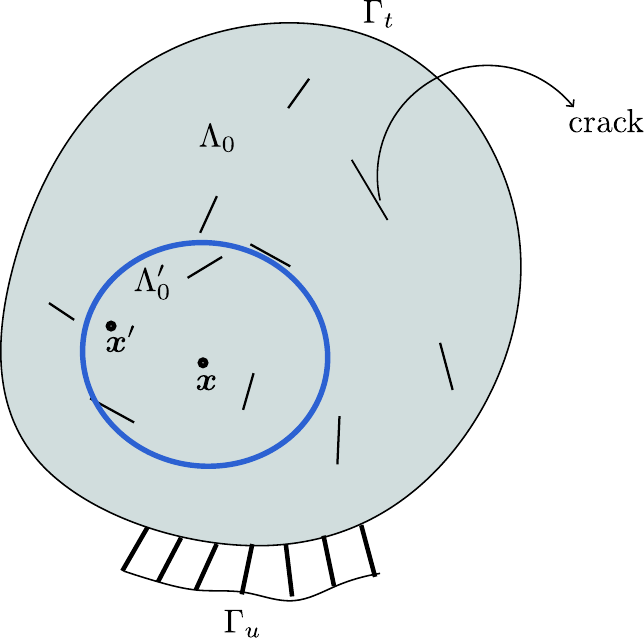}
    \caption{Deformable body $\Lambda_0$ with cracks. For the given material point $\boldsymbol{x}$, we use a subdomain $\Lambda_0'$ to define the non-local strain; $\boldsymbol{x}'$ represents a material point within $\Lambda_0'$. The boundary of the deformable body consist of the Dirichlet boundary $\Gamma_u$ and the traction boundary $\Gamma_t$.}
    \label{fig:DomainWithCracks}
\end{figure}

Let $\Lambda_0$ denote the domain with cracks as illustrated in Fig. \ref{fig:DomainWithCracks}. The boundary $\Gamma$ of  $\Lambda_0$ is partitioned into non-overlapping subsets $\Gamma_t$ and $\Gamma_u$. While external traction $\boldsymbol{t}$ is specified on $\Gamma_t$, displacement $\tilde{\boldsymbol{u}}$ is prescribed on $\Gamma_u$. The domain is assumed to undergo small strains (even though an extension to the finite deformation setting poses no complications). Given a material point $\boldsymbol{x}$, the small strain tensor $\boldsymbol{\epsilon}$ as a function of the displacement vector $\boldsymbol{u}$ is given as,
\begin{equation}\label{eq:StrainDispRel}
    \boldsymbol{\epsilon} = \frac{1}{2}\left(\nabla \boldsymbol{u} + \nabla \boldsymbol{u}^T\right)
\end{equation}
In line with our worldview of diffusing microcracks, we define a (conditional) mean strain, averaged using the transition density $g$ over a (spherical) region $\Lambda_0' \subset \Lambda_0 $ centered at $\boldsymbol{x}$ such that $g$ is negligibly small in $\Lambda_0\backslash \Lambda_0'$: 
\begin{equation}
   \boldsymbol{\epsilon}^\ast = \mathscr{R}(\boldsymbol{\epsilon})
   \label{eq:NonLocalStrain}
\end{equation}
where $ \mathscr{R}(\boldsymbol{\epsilon})$ is an integral operator defined in Eq. \eqref{AveragingOp}. Let us call the radius of $\Lambda_0'$ the effective radius. Since a random decrease in macroscopic bonds around $\boldsymbol{x}$ renders the strain at that point random, we consider the mean $\boldsymbol{\epsilon}^\ast$ a more appropriate notion of strain. Further, we define the rate of deformation tensor as $\boldsymbol{d} := \dot{\boldsymbol{\epsilon}} = \frac{1}{2}(\nabla \boldsymbol{v} +\nabla \boldsymbol{v}^T)$ where $\boldsymbol{v}$ is the velocity of the material point. The rate of non-local strain may also be determined as follows:

\begin{equation}
   \dot{\boldsymbol{\epsilon}}^\ast = \int_{\Lambda_0'} \dot{\boldsymbol{\epsilon}}g\left(\lvert\boldsymbol{x}'-\boldsymbol{x}\rvert\right)  d \Lambda_0' + \int_{\Lambda_0'} \boldsymbol{\epsilon}\dot{g}\left(\lvert\boldsymbol{x}'-\boldsymbol{x}\rvert\right)  d \Lambda_0' = \mathscr{R}(\dot{\boldsymbol{\epsilon}}) + \int_{\Lambda_0'} \boldsymbol{\epsilon}\dot{g}\left(\lvert\boldsymbol{x}'-\boldsymbol{x}\rvert\right)  d \Lambda_0'
   \label{eq:NonLocalStrainrate}
\end{equation}

\subsection{Balance of linear momentum}

Damage evolution in the deforming body is a thermodynamic process. The strain $\boldsymbol{\epsilon}$ and $g$ appear as independent state variables. $g$ may be looked upon as a damage variable whose evolution is given by Eq. \eqref{eq:phitransition}.
We use the principle of virtual power to derive the macroforce balance (i.e. the balance of linear momentum) and microforce balance (for damage evolution). Denote the external traction as $\boldsymbol{t}$, the body force as $\boldsymbol{b}$ and the virtual velocity as $\delta{\boldsymbol{v}}$. The external virtual power may be expressed as:
\begin{equation}
    \delta{P_\text{ext}} = \int_{\Lambda_0} \boldsymbol{b} \cdot \delta{\boldsymbol{v}} d\Lambda_0 + \int_{\Gamma_t} \boldsymbol{t} \cdot \delta{\boldsymbol{v}} d\Gamma_t 
    \label{Eq:ExtPower}
\end{equation}
An expression for the internal virtual power $P_{int}$, averaged using the transition density $g$, goes as follows: 

\begin{equation}\label{Eq:IntPower1}
    \begin{split}
    P_\text{int} & = \int_{\Lambda_0}  \boldsymbol{\sigma}  : \dot{\boldsymbol{\epsilon}}^\ast d \Lambda_0 + \int_{\Lambda_0} \int_{\Lambda_0'} \chi(\boldsymbol{x}') \dot{g}\left(\lvert\boldsymbol{x}'-\boldsymbol{x}\rvert\right)   d \Lambda_0' d\Lambda_0 \\ &= \int_{\Lambda_0}  \boldsymbol{\sigma}  : \mathscr{R}(\dot{\boldsymbol{\epsilon}}) d \Lambda_0 + \int_{\Lambda_0} \int_{\Lambda_0'}  \boldsymbol{\sigma} : \boldsymbol{\epsilon}(\boldsymbol{x}') \dot{g}\left(\lvert\boldsymbol{x}'-\boldsymbol{x}\rvert\right)  d \Lambda_0'  d \Lambda_0 + \int_{\Lambda_0} \int_{\Lambda_0'} \chi(\boldsymbol{x}')  \dot{g}\left(\lvert\boldsymbol{x}'-\boldsymbol{x}\rvert\right) d \Lambda_0' d\Lambda_0    \end{split}
\end{equation}
where we have used Eq. \eqref{eq:NonLocalStrainrate}. $\chi$ is a microstress conjugate to $\dot{g}$.

$\mathcal{R}$ being self-adjoint (see Eq. (5) in \cite{polizzotto2001nonlocal} and references therein), Eq. \eqref{Eq:IntPower1} may be written as:

\begin{equation}
    P_\text{int} = \int_{\Lambda_0}  \mathscr{R}(\boldsymbol{\sigma}) : \dot{\boldsymbol{\epsilon}} d \Lambda_0 +\int_{\Lambda_0} \int_{\Lambda_0'}  \boldsymbol{\sigma} : \boldsymbol{\epsilon}(\boldsymbol{x}') \dot{g}\left(\lvert\boldsymbol{x}'-\boldsymbol{x}\rvert\right) d \Lambda_0'  d \Lambda_0+ \int_{\Lambda_0} \int_{\Lambda_0'} \chi(\boldsymbol{x}')  \dot{g}\left(\lvert\boldsymbol{x}'-\boldsymbol{x}\rvert\right) d \Lambda_0' d\Lambda_0 
    \label{Eq:IntPower2}
\end{equation}
We denote $\mathscr{R}(\boldsymbol{\sigma})$ as $\boldsymbol{\sigma}^\ast$.
Virtual variation of the internal power is thus given by: 

\begin{equation}
\delta{P_\text{int}} = \int_{\Lambda_0} \mathscr{R}(\boldsymbol{\sigma}) : \delta{\dot{\boldsymbol{\epsilon}}} d \Lambda_0 + \int_{\Lambda_0} \int_{\Lambda_0'}  \boldsymbol{\sigma} : \boldsymbol{\epsilon}(\boldsymbol{x}') \delta{\dot{g}}\left(\lvert\boldsymbol{x}'-\boldsymbol{x}\rvert\right)  d \Lambda_0'  d \Lambda_0 + \int_{\Lambda_0} \int_{\Lambda_0'} \chi(\boldsymbol{x}') \delta{\dot{g}}\left(\lvert\boldsymbol{x}'-\boldsymbol{x}\rvert\right) d \Lambda_0' d\Lambda_0 
    \label{Eq:IntPower3}
\end{equation}
$\boldsymbol{\sigma}^\ast$ is symmetric by prescription and hence satisfies the balance of angular momentum. Using the divergence theorem and $\dot{\boldsymbol{\epsilon}} = \boldsymbol{d}$, we may write   as  
\begin{equation}
  \int_{\Lambda_0} \boldsymbol{\sigma}^\ast :\delta{\boldsymbol{d}}  d\Lambda_0 = \int_{\Lambda_0} \boldsymbol{\sigma}^\ast :\nabla \delta{\boldsymbol{v}}  d\Lambda_0
    = \int_{\Gamma_t} ( \boldsymbol{\sigma}^\ast\boldsymbol{n})\cdot \delta{\boldsymbol{v}} d\Gamma_t - \int_{\Lambda_0} (\nabla\cdot \boldsymbol{\sigma}^\ast)\cdot \delta{\boldsymbol{v}}d\Lambda_0 
    \label{Eq:DivTheomStress}
\end{equation}
where $\boldsymbol{n}$ is the unit outward normal to $\Gamma_t$.
The macroforce and microforce balances can be derived by applying the principle of virtual power, viz.  
$\delta{P_\text{ext}} = \delta{P_\text{int}}$. In addition, we assume $\boldsymbol{b}=\boldsymbol{0}$ and use Eqs. \eqref{Eq:ExtPower}, \eqref{Eq:IntPower3} and \eqref{Eq:DivTheomStress} to get the local macroforce balance as:
\begin{subequations}\label{macroforcebalance}
\begin{align}
    \nabla \cdot\boldsymbol{\sigma}^\ast &= 0\,\,\,\, \text{on $\Lambda_0$}\\
   \boldsymbol{\sigma}^\ast \boldsymbol{n} &= \boldsymbol{t} \,\,\,\, \text{on $\Gamma_t$}
\end{align}    
\end{subequations}
The local microforce balance is of the following form: 
\begin{equation}\label{microforcebalance}
\begin{split}
\chi + \boldsymbol{\sigma} : \boldsymbol{\epsilon} =0\,\,\,\, \text{on $\Lambda_0$}
\end{split}    
\end{equation}

The above framework shows that our worldview aligns well with certain well known continuum damage models. However, as we demonstrate below, the microforce balance  \mbox{\eqref{microforcebalance}} is inherently incorporated into the constitutive equation for $g$, obviating the need for its solution via expensive numerical integration. 
\subsection{Constitutive equations and dissipation inequality}

Let the free energy density $\psi$ depend on $\boldsymbol{\epsilon}$ and $g$ as: 
\begin{equation}\label{eq:psiform}
\psi=\hat{\psi}(\boldsymbol{\epsilon},\boldsymbol{\epsilon}^\ast(\boldsymbol{\epsilon},g))
\end{equation}
This choice for $\psi$ is motivated by the existing literature on nonlocal elasticity (see  \mbox{\cite{polizzotto2001nonlocal}} and the references therein), even though other choices, such as $\psi=\hat{\psi}(\boldsymbol{\epsilon}^\ast)$, are certainly possible. Note that, given the material point $\boldsymbol{x}$, the strain $\boldsymbol{\epsilon}=\boldsymbol{\epsilon} (\boldsymbol{x})$ is non-random. We anticipate that these alternatives would exhibit a similar behavior, possibly with modified values of the material parameters.

Evaluating the material time derivative of Eq. \eqref{eq:psiform}, we have:
\begin{equation}\label{Eq: mat_time_der}
\begin{split}
\int_{\Lambda_0} \Dot{\hat{\psi}}(\boldsymbol{\epsilon},\boldsymbol{\epsilon}^\ast) d{\Lambda_0} 
&=  \int_{\Lambda_0}\frac{\partial \hat{\psi}}{\partial \boldsymbol{\epsilon}}: \Dot{\boldsymbol{\epsilon}}d{\Lambda_0} + \int_{\Lambda_0}\int_{\Lambda_0'} \frac{\partial\hat{\psi}}{\partial \boldsymbol{\epsilon}^\ast} : \Dot{\boldsymbol{\epsilon}}(\boldsymbol{x}')g(\lvert\boldsymbol{x}'-\boldsymbol{x}\rvert)d\Lambda_0' 
d\Lambda_0 \\&+ \int_{\Lambda_0}\int_{\Lambda_0'} \frac{\partial \hat{\psi}}{\partial \boldsymbol{\epsilon}^\ast} : \boldsymbol{\epsilon}(\boldsymbol{x}')\Dot{g}(\lvert\boldsymbol{x}'-\boldsymbol{x}\rvert)d\Lambda_0' 
d\Lambda_0 \\
&=     \int_{\Lambda_0}\frac{\partial \hat{\psi}}{\partial \boldsymbol{\epsilon}}: \Dot{\boldsymbol{\epsilon}}d{\Lambda_0} + \int_{\Lambda_0}\frac{\partial \hat{\psi}}{\partial \boldsymbol{\epsilon}^\ast} : \mathcal{R}(\Dot{\boldsymbol{\epsilon}}) d{\Lambda_0} + \int_{\Lambda_0}\int_{\Lambda_0'} \frac{\partial \hat{\psi}}{\partial \boldsymbol{\epsilon}^\ast} : \boldsymbol{\epsilon}(\boldsymbol{x}')\Dot{g}(\lvert\boldsymbol{x}'-\boldsymbol{x}\rvert)d\Lambda_0' 
d\Lambda_0     
\end{split}
\end{equation}
where Eq. \eqref{eq:NonLocalStrainrate} has been used to resolve $\dot{ \boldsymbol{\epsilon}}^\ast$. Since $\mathcal{R}$ is self-adjoint, Eq. \eqref{Eq: mat_time_der} could be rewritten as:
\begin{equation}\label{Eq: mat_time_der_final}
    \begin{split}
 \int_{\Lambda_0} \Dot{\psi}(\boldsymbol{\epsilon}, \boldsymbol{\epsilon}^\ast) d{\Lambda_0} &=     \int_{\Lambda_0}\frac{\partial \hat{\psi}}{\partial \boldsymbol{\epsilon}}: \Dot{\boldsymbol{\epsilon}}d{\Lambda_0} + \int_{\Lambda_0} \mathscr{R}(\frac{\partial \hat{\psi}}{\partial \boldsymbol{\epsilon}^\ast}) : \Dot{\boldsymbol{\epsilon}} d{\Lambda_0} + \int_{\Lambda_0}\int_{\Lambda_0'} \frac{\partial \hat{\psi}}{\partial \boldsymbol{\epsilon}^\ast} : \boldsymbol{\epsilon}(\boldsymbol{x}')\Dot{g}(\lvert\boldsymbol{x}'-\boldsymbol{x}\rvert)d\Lambda_0' 
d\Lambda_0\\
  &=     \int_{\Lambda_0} \left(\frac{\partial \hat{\psi}}{\partial \boldsymbol{\epsilon}}  +\mathscr{R}(\frac{\partial \hat{\psi}}{\partial \boldsymbol{\epsilon}^\ast}) \right) : \Dot{\boldsymbol{\epsilon}} d{\Lambda_0} + \int_{\Lambda_0}\int_{\Lambda_0'}\frac{\partial \hat{\psi}}{\partial \boldsymbol{\epsilon}^\ast} : \boldsymbol{\epsilon}(\boldsymbol{x}')\Dot{g}(\lvert\boldsymbol{x}'-\boldsymbol{x}\rvert)d\Lambda_0' 
d\Lambda_0
\end{split}       
\end{equation}

Consistent with the second law of thermodynamics, the energy imbalance for a purely mechanical system, such as ours, is given by:
\begin{equation}\label{eq:energy_imbalance}
\begin{split}
    &P_\text{int} - \int_{\Lambda_0}\Dot{\psi} d\Lambda_0 \ge 0
    \end{split}
    \end{equation}
We may determine the constitutive restrictions on the thermodynamic forces through the Coleman-Noll procedure. Rewriting the inequality by incorporating the expression of $P_\text{int}$ from Eq. $\eqref{Eq:IntPower2}$, we have: 
\begin{equation}\label{eq:energy_imbalance2}
\begin{split}
 \int_{\Lambda_0}  \mathscr{R}(\boldsymbol{\sigma}) : \dot{\boldsymbol{\epsilon}} d \Lambda_0 +\int_{\Lambda_0} \int_{\Lambda_0'}  (\boldsymbol{\sigma} : \boldsymbol{\epsilon}+\chi)  \dot{g}\left(\lvert\boldsymbol{x}'-\boldsymbol{x}\rvert\right) d \Lambda_0'  d \Lambda_0  - \int_{\Lambda_0}\Dot{\psi} d\Lambda_0  \ge 0
\end{split}
\end{equation}
Using the microforce balance Eq. $\eqref{microforcebalance}$ in the inequality, we obtain:

\begin{equation}\label{eq:energy_imbalance3}
\int_{\Lambda_0}\boldsymbol{\sigma}^\ast(\boldsymbol{x}) : \dot{\boldsymbol{\epsilon}}(\boldsymbol{x})  d\Lambda_0 - \int_{\Lambda_0}\Dot{\psi} d\Lambda_0 \ge 0
\end{equation}
Substitution of Eq. \eqref{Eq: mat_time_der_final} in the energy imbalance in the global form, we arrive at:
\begin{equation}
    \int_{\Lambda_0} \left(\boldsymbol{\sigma}^\ast - \left( \frac{\partial \hat\psi}{\partial \boldsymbol{\epsilon}} +\mathscr{R}(\frac{\partial \hat\psi}{\partial \boldsymbol{\epsilon}^\ast})\right)\right):\Dot{\boldsymbol{\epsilon}} d\Lambda_0 - \int_{\Lambda_0}\int_{\Lambda_0'}\frac{\partial \hat\psi}{\partial \boldsymbol{\epsilon}^\ast} : \boldsymbol{\epsilon}(\boldsymbol{x}')\Dot{g}(\lvert\boldsymbol{x}'-\boldsymbol{x}\rvert)d\Lambda_0' 
d\Lambda_0\ge 0 
\end{equation}
$\boldsymbol{\sigma}$ must have the following form to satisfy the energy imbalance.
\begin{equation}\label{eq:sigma_ast}
       \boldsymbol{\sigma}^\ast  = \frac{\partial \hat\psi}{\partial \boldsymbol{\epsilon}} +\mathscr{R}(\frac{\partial \hat\psi}{\partial \boldsymbol{\epsilon}^\ast})\\
\end{equation}
With the restriction on $\boldsymbol{\sigma}$ so established, the restriction on $\chi$ also follows. The reduced energy imbalance or the dissipation inequality is given by:
\begin{equation}
\frac{\partial \hat\psi}{\partial \boldsymbol{\epsilon}^\ast} : \int_{\Lambda_0'}\boldsymbol{\epsilon}(\boldsymbol{x}')\Dot{g}(\lvert\boldsymbol{x}'-\boldsymbol{x})\rvert d\Lambda_0' \le 0
\label{eq:DissiPainEq1}
\end{equation}
The inequality can be ensured by a proper evolution of $g$ and hence by a suitable choice of the killing rate $G$. Note that constitutive Eq. \mbox{\eqref{eq:sigma_ast}} arises upon the incorporation of the balance law \mbox{\eqref{microforcebalance}} into the inequality \mbox{\eqref{eq:energy_imbalance2}}. Consequently, when solving the governing equations, it is only necessary to solve the linear momentum balance Eq. \mbox{\eqref{macroforcebalance}}. Damage evolution may then be determined from Eq. \mbox{\eqref{eq:phitransition}}.
\subsection{Specialized constitutive theory}
\label{sec:SpecializatioofConstitutiveFunction}
In this section, we describe the specific form of the killing rate $G$ and that of stress via an appropriate expression of the free energy density $\psi$.

\subsubsection{Structure of the killing rate \texorpdfstring{$G$}{G}}
The form for $G$ must ensure non-violation of the dissipation inequality i.e. $\dot{g}\le 0$, so that $G$ must remain positive (almost surely). Furthermore, to ensure physically meaningful damage evolution, the following items need to be introduced.
\begin{itemize}
    \item The existence of a (scalar measure of) critical stress below which the response is purely elastic; we denote it by $\sigma_{\text{crit}}$
    \item A parameter $G_0$ controlling the rate of damage evolution
    \item A term that vanishes with unloading to ensure no change in the state of damage (during unloading)
\end{itemize}
Based on the three conditions stipulated above, we assume the following form of $G$:
\begin{equation}\label{eq:KillingRate}
    G = G_0 \langle \frac{\sigma_\text{eq}}{\sigma_{\text{crit}}}-1\rangle_+ \langle\Dot{\sigma}_\text{eq}\rangle_+
\end{equation}
where $\langle \bullet\rangle_+$is defined as $ \frac{\bullet + \lvert\bullet\rvert}{2}$. Note that the material behaves elastically till the critical stress $\sigma_{\text{crit}}$ is reached. We interpret  $\sigma_{\text{crit}}$ as the uniaxial tensile strength, which may be determined experimentally. $G_0$ is the fracturing rate, and its value for the material and loading condition can be calibrated based on post-elastic response. A similar form of the driving term has been considered using strain normal to the crack plane  \mbox{\cite{thamburaja2021fracture}}. This ensures physically meaningful damage evolution as it captures typical experimental features of load-displacement curves, such as the initial elastic limit, pre-peak non-linear response, post-peak softening and irreversibility of damage during unloading.
The equivalent stress, $\sigma_{\text{eq}}$, is calculated as: 
\begin{equation}\label{eq:EqStress}
    \sigma_{\text{eq}} = \sqrt{2\hat{\psi}^{+}E}
\end{equation}
where $E$ is the Young's modulus and $\hat{\psi}^{+}$, the part of $\hat{\psi}$ contributing to damage. Use of the equivalent stress here is similar to the approach in \mbox{\cite{lemaitre1985continuous}}, wherein damage evolution is governed by a quantity similar to strain energy density, and $\sigma_{\text{eq}}$ is used as a criterion for the evolution (similar to yield stress in plasticity). It may be used to represent the elastic limit in our case when $\sigma_{\text{eq}}$ reaches $\sigma_\text{crit}$. 

\subsubsection{Energy density}
We adopt the following form of $\psi$ from \cite{polizzotto2001nonlocal}:
\begin{equation}\label{eq:Energy_density}
\hat{\psi}(\boldsymbol{\epsilon}, \boldsymbol{\epsilon}^\ast) =  \frac{\lambda}{2}(\text{tr}\boldsymbol{\epsilon})(\text{tr}\boldsymbol{\epsilon}^\ast)+\mu \boldsymbol{\epsilon}:\boldsymbol{\epsilon}^{\ast}  \end{equation}
Substituting Eq. \eqref{eq:sigma_ast} in Eq. \eqref{eq:Energy_density}, we obtain:
\begin{equation}
\boldsymbol{\sigma}^\ast = \left(\frac{1}{2}\lambda(\text{tr}\boldsymbol{\epsilon}^\ast)+\mu \boldsymbol{\epsilon}^{\ast}\right) + \left(\frac{1}{2}\lambda(\text{tr}\boldsymbol{\epsilon}^\ast)+\mu \boldsymbol{\epsilon}^{\ast}\right)
= \lambda(\text{tr}\boldsymbol{\epsilon}^\ast)+2\mu \boldsymbol{\epsilon}^{\ast}
\end{equation}\label{detailedexpressionstress}
Various methods are available to define $\hat{\psi}^{+}$, each with its limitations and advantages (see \cite{ziaei2023orthogonal} and references therein). For this work, we choose the spectral decomposition proposed by \cite{miehe2010thermodynamically}, and thus define $\hat{\psi}^+$ as:
\begin{equation}
\hat{\psi}^+(\boldsymbol{\epsilon}, \boldsymbol{\epsilon}^\ast) =    \frac{\lambda}{2}\langle \text{tr}\boldsymbol{\epsilon}\rangle_+\langle\text{tr}\boldsymbol{\epsilon}^\ast\rangle_+ + \mu \boldsymbol{\epsilon}^+:\boldsymbol{\epsilon}^{\ast^+}\end{equation}
where
\begin{subequations}
\begin{align}
    \boldsymbol{\epsilon}^{+} = \sum_{i=1}^3 \langle\epsilon_i \rangle_+\boldsymbol{p}_i\otimes\boldsymbol{p}_i,\\   \boldsymbol{\epsilon}^{\ast^+} = \sum_{i=1}^3 \langle\epsilon_i^{\ast}\rangle_+ \boldsymbol{p}_i^{\ast}\otimes\boldsymbol{p}_i^{\ast}    \end{align}
\end{subequations}
$\epsilon_i$ and $\epsilon_i^\ast$ $(i = 1,2,3)$ are the eigenvalues and $p_i$, $p_i^\ast$ are the eigenvectors of $\boldsymbol{\epsilon}$ and $\boldsymbol{\epsilon}^\ast$, respectively. $\lambda$ and $\mu$ are Lame's parameters. For numerical illustrations, elastic constants such as Young's modulus ($E$) and Poisson's ratio ($\nu$) have been taken from reported experiments.

\begin{figure}[ht!]
    \centering
    \begin{subfigure}[b]{0.48\textwidth}
        \centering
        \includegraphics[width=0.95\textwidth]{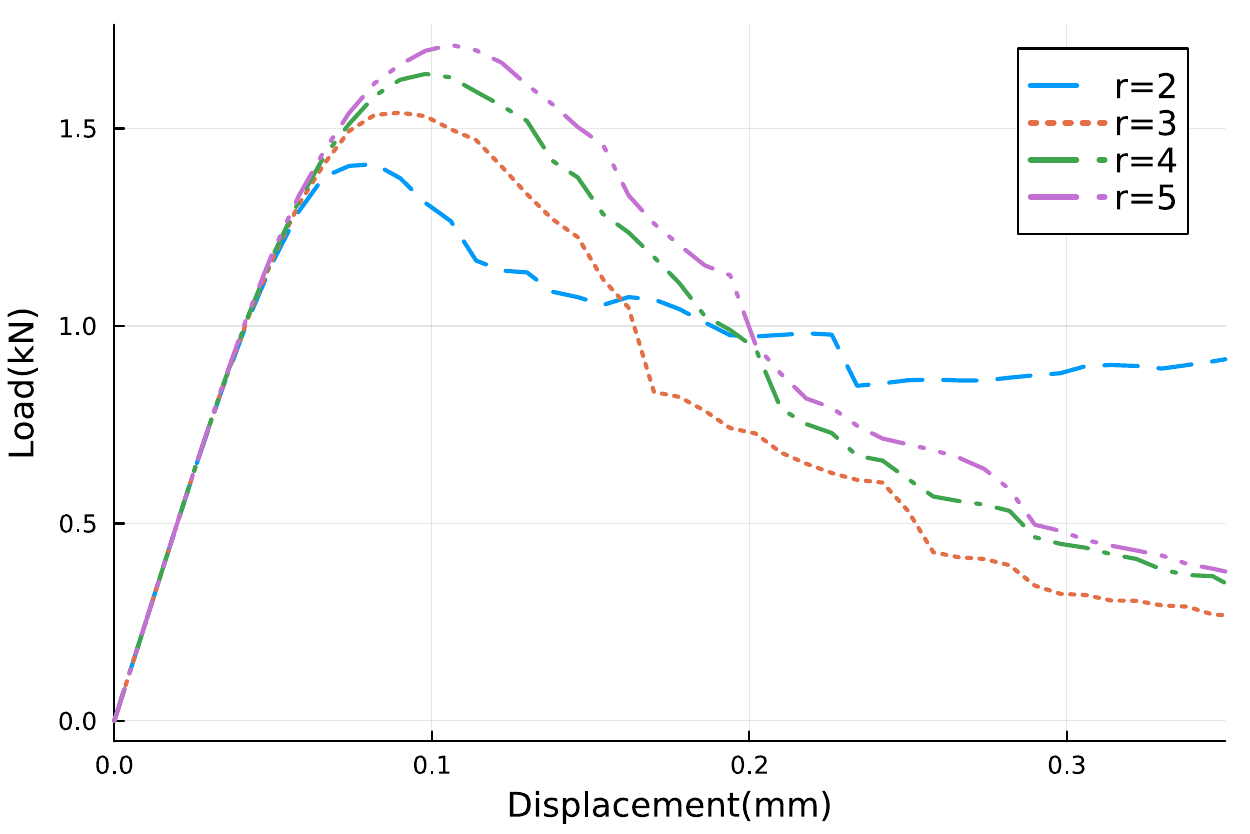}
        \caption{}
        \label{fig:threePB_r_variation}
    \end{subfigure}
    \hfill
    \begin{subfigure}[b]{0.48\textwidth}
        \centering
        \includegraphics[width=0.95\textwidth]{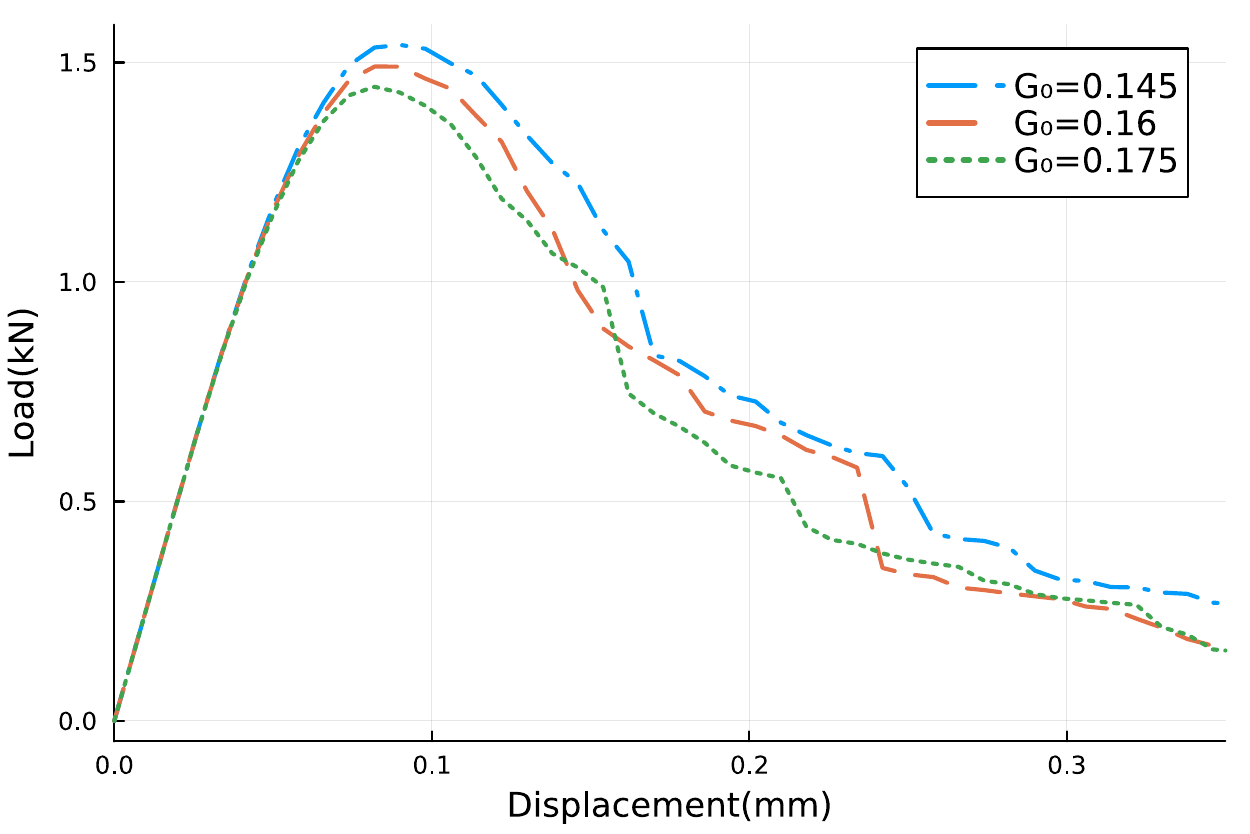}
        \caption{}
        \label{fig:threePEffectofG0ft2p4}
    \end{subfigure}
\caption{(a) Effect of variable effective radii for a fixed average element size of $1.5$ mm; (b) Effect of $G_0$ for three-point bending test with $\sigma_{\text{crit}} = 2.4$ MPa}
    \label{fig:threePB_combined}
\end{figure}

\SetKwInput{KwInput}{Input} 
\begin{algorithm}[H]
\label{algo:Algo1}
\DontPrintSemicolon
\KwInput{Geometric and material properties}
\caption{Recursive implementation of FeynKrack in $[t_n,t_{n+1}]$}

Initialization. $\boldsymbol{u}_\text{n}$, $G_\text{n}$ and $g_\text{n}$ are known at time $t_\text{n}$. Update prescribed load or displacement at the current time $t_\text{n+1}$.\\
\For{i = 1: MaxIter} {Determine $\boldsymbol{u}_{\text{n+1}}^i$ by solving the linear momentum balance equation. \tcp*{Eq.\eqref{macroforcebalance}}

Compute strains using the strain-displacement relation. \tcp*{Eq.\eqref{eq:StrainDispRel}}

Compute the equivalent stress. \tcp*{Eq.\eqref{eq:EqStress}}

Compute $G_{\text{n+1}}^i$.  \tcp*{Eq.\eqref{eq:KillingRate}}

Compute $g_{\text{n+1}}^i$. \tcp*{Eq.\eqref{eq:phitransition}}

Compute \textit{error} = $\lvert\boldsymbol{u}_{\text{n+1}}^i-\boldsymbol{u}_{\text{n+1}}^{i-1}\rvert$ \tcp* {user dependent}
\If{error $ \le $ tol}{    $\boldsymbol{u}_{\text{n+1}}$ = $\boldsymbol{u}_{\text{n+1}}^i$;
    $G_{\text{n+1}}$ = $G_{\text{n+1}}^i$; 
    $g_{\text{n+1}}$ = $g_{\text{n+1}}^i$\\
    break the loop
   }
$\boldsymbol{u}_{\text{n+1}}$ = $\boldsymbol{u}_{\text{n+1}}^\textit{MaxIter}$;
$G_{\text{n+1}}$ = $G_{\text{n+1}}^\textit{MaxIter}$; 
$g_{\text{n+1}}$ = $g_{\text{n+1}}^\textit{MaxIter}$}
\end{algorithm}
\section{FeynKrack: Numerical results}
\label{sec:Validation}

We have solved Eq. \eqref{macroforcebalance}, the macroforce balance, using the FEM implemented via the Gridap open source package \cite{verdugo2022software}. Eq. \mbox{\eqref{eq:phitransition}} is used to determine the extent of damage at a particular loading step. Although the closed-form expression is valid only for an infinite domain, such an approximation makes sense even for finite domains when considering small diffusion coefficient and time-increment, so that variance ($\beta^2 \Delta t$) of the Brownian increment remains small. Techniques like Doob's h-transform \cite{roy2017stochastic} may also be used to develop a more accurate theory for finite domains with boundaries; this could be explored in future. The integral of Eq. \mbox{\eqref{eq:phitransition}} has been calculated at each $\boldsymbol{x}$ to plot the damage contour. Previous studies have shown that CDM exhibits mesh-sensitive behavior. To mitigate this issue, it has been suggested that the driving term be averaged over a small neighborhood \mbox{\cite{pijaudier1987nonlocal}}. This insight guides our choice of the optimal effective radius, denoted as $r$. FeynKrack is validated through numerical simulations of several boundary value problems involving quasi-brittle fracture. Note that refined mesh in numerical simulations corresponds to the damaged parts, where nonlocal interactions are active, and sparse mesh to parts away from damage, where the classical (i.e. local) laws of mechanics are deemed to hold. Concern with the calculation of nonlocal response quantities therefore arises only in zones with refined mesh (where, depending on the refinement, more nodal points could be made available for numerical integration). Near the boundary, we assume a smaller effective radius to avoid contribution of points beyond the boundary. Since $\beta^2 \Delta t = 0$ is the degenerate case (where damage does not propagate), a very small value might however cause inaccuracy when the crack approaches the boundary. In line with non-local CDM and in cases where this inaccuracy is of concern, one way is a fictitious extension of the physical boundary \mbox{\cite{chen2015selecting}}. Let the damage process over the time interval $[0,T]$ be of interest. Field quantities are then calculated at the discrete time instants $0<t_1<t_2<...t_\text{n}<t_\text{n+1} <...T$. Using Algorithm \mbox{\ref{algo:Algo1}}, the unknown fields are obtained at the current time $t_\text{n+1}$ given the converged solution at $t_\text{n}$. The first set of simulations comprises of three-point bending tests on concrete specimens and these are used to study the influence of various parameters such as fracturing rate ($G_0$), effective radius ($r$) etc. For validation purposes, experimental results from the literature are compared with simulations of three-point bending tests on notched concrete specimens, mixed-mode loading tests on L-shaped specimens, and uniaxial compression tests on rocks with internal inclined cracks.
The model's capabilities are then demonstrated through simulations of multi-modal fracture in three dimensions, focusing on the inclined crack problem and the Brazilian test. The codes used in generating results for both 2D and 3D can be accessed through \mbox{\href{https://github.com/Debasish-Roy-IISc/FeynKrack.git}{codes for FeynKrack}}. \hl{For 2D problems, we use constant-strain triangles and assume  plane stress conditions. 3D problems are discretized with tetrahedral elements.}

\begin{figure}[ht!]
    \centering
    \begin{subfigure}[b]{0.7\textwidth}
         \centering
\includegraphics[width=0.6\textwidth]{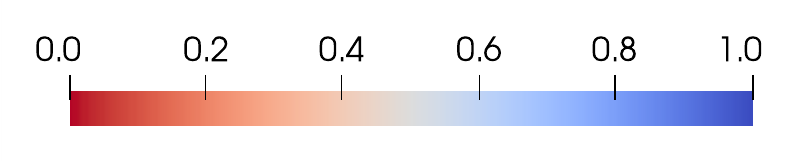}
    \label{fig:colorbar_hor}
     \end{subfigure}
 \centering
    \begin{subfigure}[b]{0.32\textwidth}
         \centering
\includegraphics[width=1.0\textwidth]{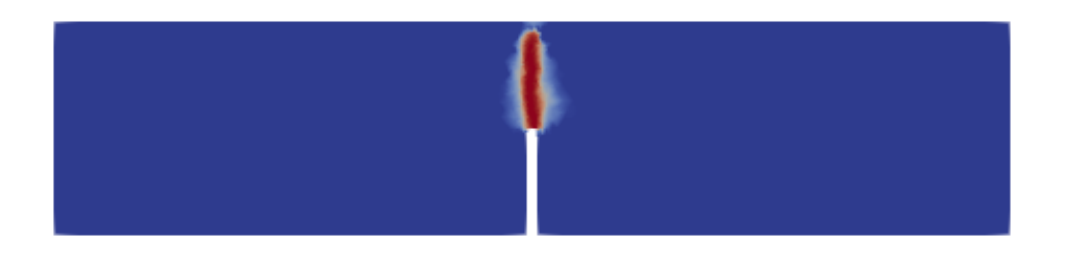}
    \caption{}
    \label{fig:3pBendmsh1r2}
     \end{subfigure}
     \hfill
    \begin{subfigure}[b]{0.32\textwidth}
         \centering
 \includegraphics[width=1.0\textwidth]{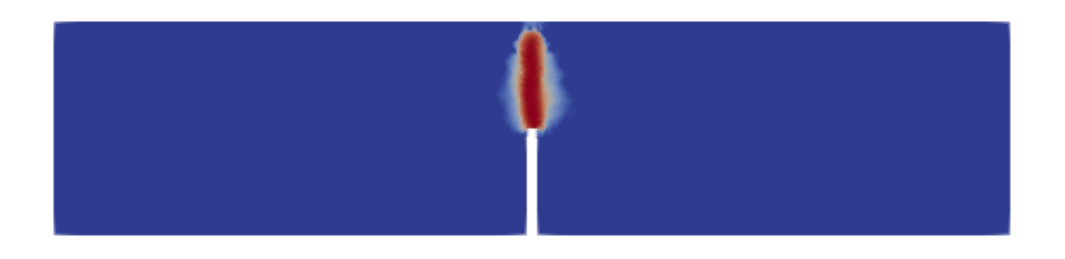}
    \caption{}
    \label{fig:3pBendmsh1r4}
     \end{subfigure}
     \hfill
\begin{subfigure}[b]{0.32\textwidth}
         \centering
 \includegraphics[width=1.0\textwidth]{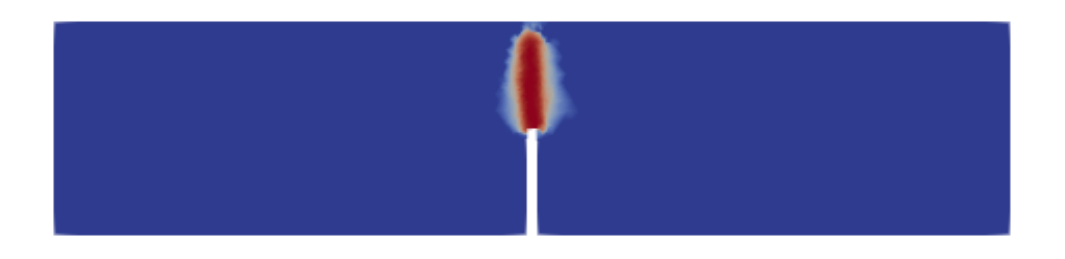}
    \caption{}
    \label{fig:3pBendmsh1r6}
     \end{subfigure}
  \caption{Crack paths in a three-point bending test; (a) $r$ = 3 (b) $r$ = 4 (c) $r$ = 5}
\label{fig:ThreePBendCrack_r_var}
\end{figure}

\subsection{Parametric studies}
We now study the effects of effective radius ($r$), fracturing rate ($G_0$), etc. through numerical simulations of three-point bending tests on plain concrete specimens \cite{kormeling1983determination}. The geometry and boundary conditions are shown in Fig. \ref{fig:ThreePointBendingGeom}. Material parameters used are as follows: $E$, Young's modulus = $20$ GPa and $\nu$, Poisson's ratio = $0.2$, and $\sigma_{\text{crit}}$, the uniaxial tensile strength = $2.4$ Mpa. The simulations are carried out using constant strain triangle elements.

\subsubsection{Effective radius \texorpdfstring{$r$}{r}}
The parameter $r$ is related to the length scale of heterogeneity (see \mbox{\cite{thamburaja2021fracture}} and the references therein). We study its effect on the response as a guide to an appropriate choice. The mesh around the notch is refined with an average element size ($m_{\text{avg}}$) 1.5 mm. The influence of $r$ on the load-displacement behavior is depicted in Fig. \ref{fig:threePB_r_variation}. When the effective radius increases, so do the peak load and the crack width (Fig. \ref{fig:ThreePBendCrack_r_var}). We have observed that an effective radius below 3 mm does not accurately represent the softening characteristics. Therefore, for the rest of the simulations, we use $m_{\text{avg}}\le\frac{r}{2}$.

\begin{figure}[htbp]
    \centering
    \begin{minipage}[t]{0.48\textwidth}
        \centering
\includegraphics[width=\textwidth,height=8cm,keepaspectratio,valign=t]{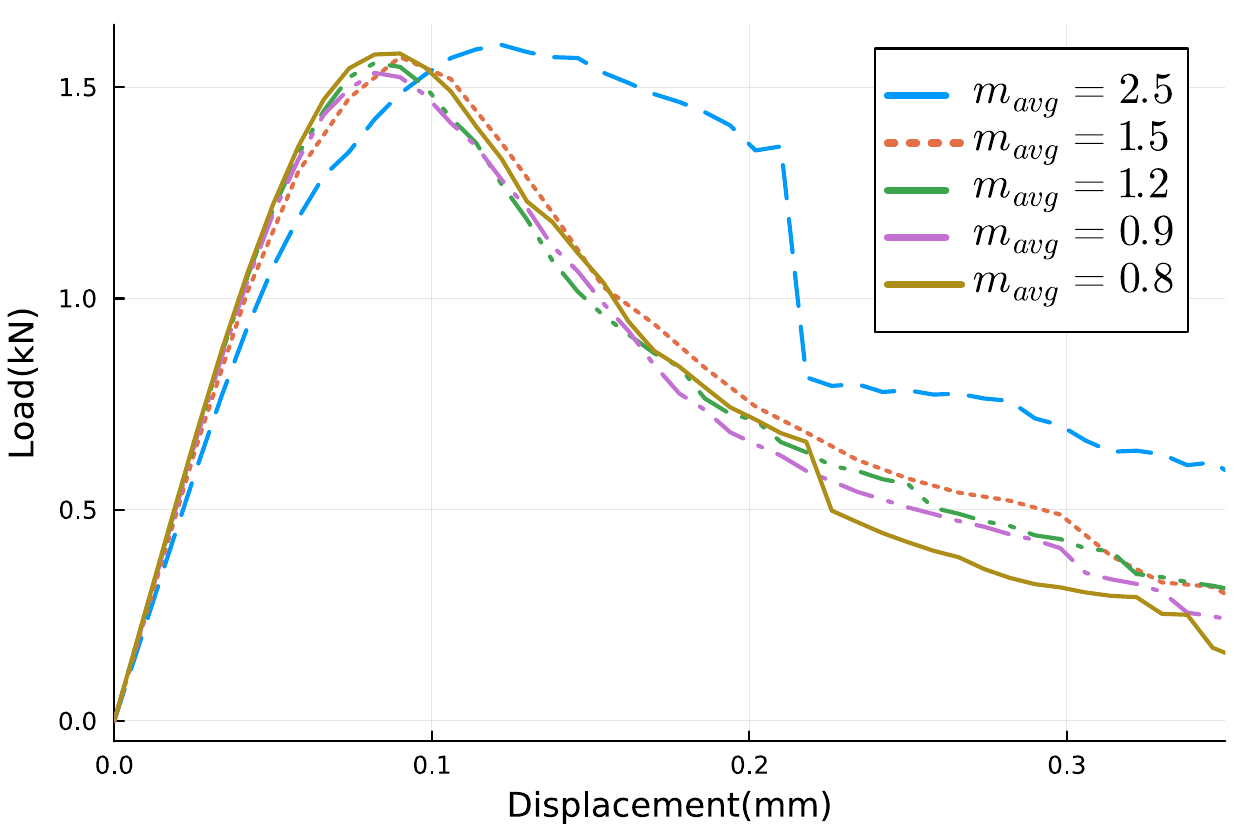}     \subcaption{}\label{fig:threePMeshConverg}
    \end{minipage}\hfill
    \begin{minipage}[t]{0.48\textwidth}
\includegraphics[width=\textwidth,height=2cm,keepaspectratio,valign=t]{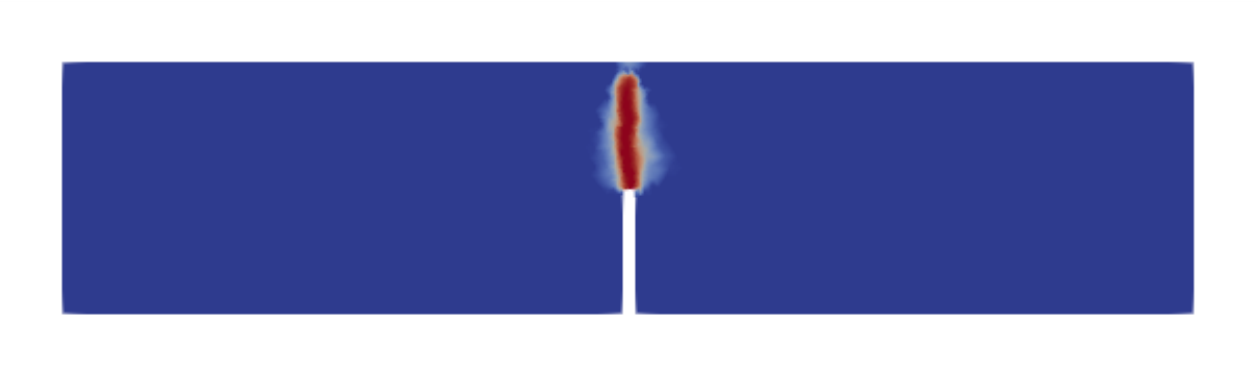}
        \subcaption{}\label{fig:3PBend_mesh_1p5}
\includegraphics[width=\textwidth,height=2cm,keepaspectratio]{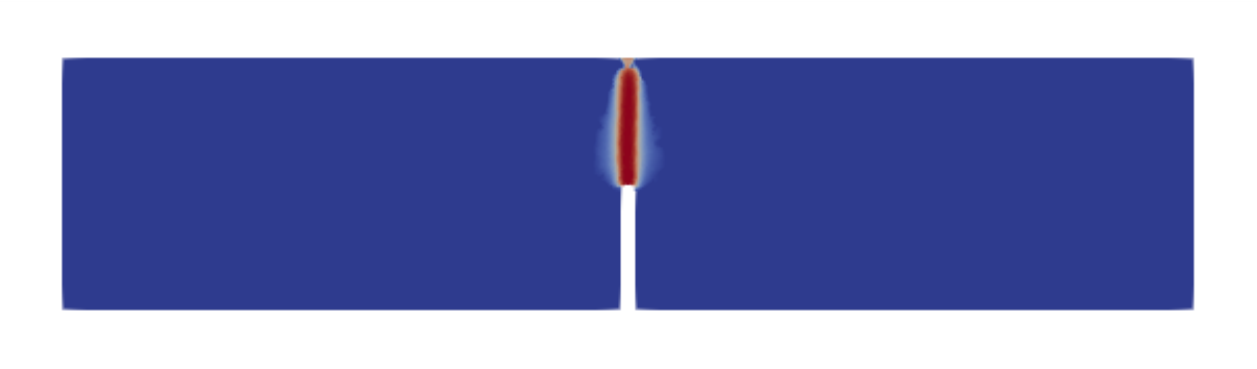}
        \subcaption{}\label{fig:3PBend_mesh_0p8}
    \end{minipage}
    \caption{(a) Mesh convergence study for three-point bending test; Crack paths in a three-point bending test for (b) $m_{\text{avg}}$ = 1.5 \hl{(c) $m_{\text{avg}}$ = 0.8} }
    \label{fig:combined}
\end{figure}
\subsubsection{Fracturing rate \texorpdfstring{$G_0$}{G_0}}
To examine the effect of the parameter $G_0$, we fix the effective radius at 3 mm whilst maintaining the other parameters as in the previous section. Fig. \ref{fig:threePEffectofG0ft2p4} shows the load-displacement curves for varying $G_0$. As anticipated, $G_0$ governs the fracture rate, so lower $G_0$ leads to higher peak loads and slower softening.
\subsection{Mesh convergence}
Holding all other material parameters the same as in the previous section, $G_0$ is taken as 0.145 mm$^2$/N, and the mesh-size near the notch is varied over a range of average element lengths ($m_{\text{avg}}$), viz. from 2.5 mm to 0.8 mm. The response is plotted in Fig. \ref{fig:threePMeshConverg}. We find that the solution converges when $m_{\text{avg}}$ = 1.5 mm. To ensure mesh objectivity, we maintain a constant ratio of 0.5 between the average element size and the effective radius throughout the rest of our study, following the guidelines in \cite{khodabakhshi2019nonlocal}. 

\begin{figure}[ht!]
\centering
    \begin{subfigure}[b]{0.68\textwidth}
         \centering
\includegraphics[width=0.5\textwidth]{colorbar_hor.pdf}
    \label{fig:colorbar_hor}
     \end{subfigure}
 \centering
 \begin{subfigure}[b]{0.32\textwidth}
         \centering
 \includegraphics[width=0.9\textwidth]{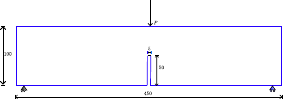}
         \caption{}
         \label{fig:ThreePointBendingGeom}
     \end{subfigure}
    \begin{subfigure}[b]{0.32\textwidth}
         \centering
 \includegraphics[width=\textwidth]{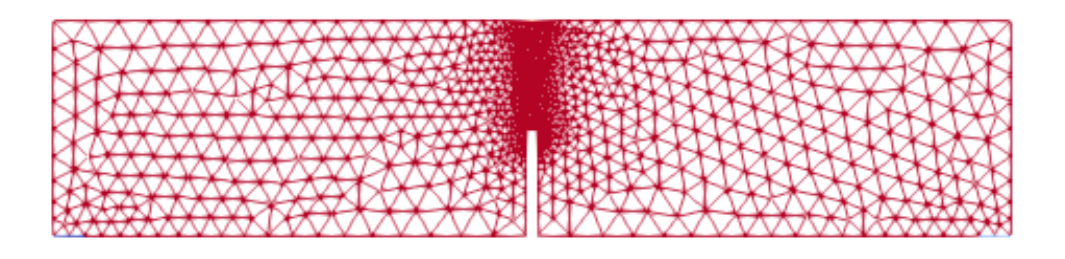}
         \caption{}
         \label{fig:ThreePBendMesh_o16_h1p5}
     \end{subfigure}
    \begin{subfigure}[b]{0.32\textwidth}
         \centering
 \includegraphics[width=1\textwidth]{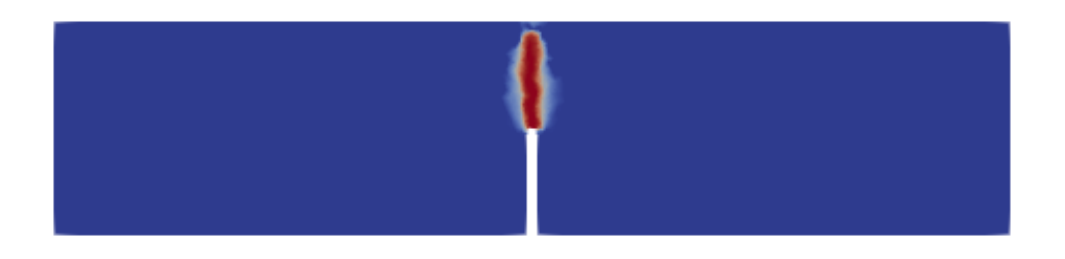}
         \caption{}
          \label{fig:ThreePBendCrack_Exp_comp}
     \end{subfigure}
  \caption{(a) Geometry and boundary conditions for the three-point bending specimen (all dimensions are in mm) (b) Three-point bending test on a notched concrete beam: FE mesh with $m_{\text{avg}}$ = 1.5 mm around the centroid of the beam and $m_{\text{avg}}$ = 16 mm in the rest of the domain; (c) contour plot of damaged surface}
  \label{fig:ThreePBendDamSurf}
\end{figure}
\subsection{Three point bending test}
\label{sec:ThreePBend}
\begin{figure}[ht!]
    \centering
\includegraphics[width=0.6\textwidth]{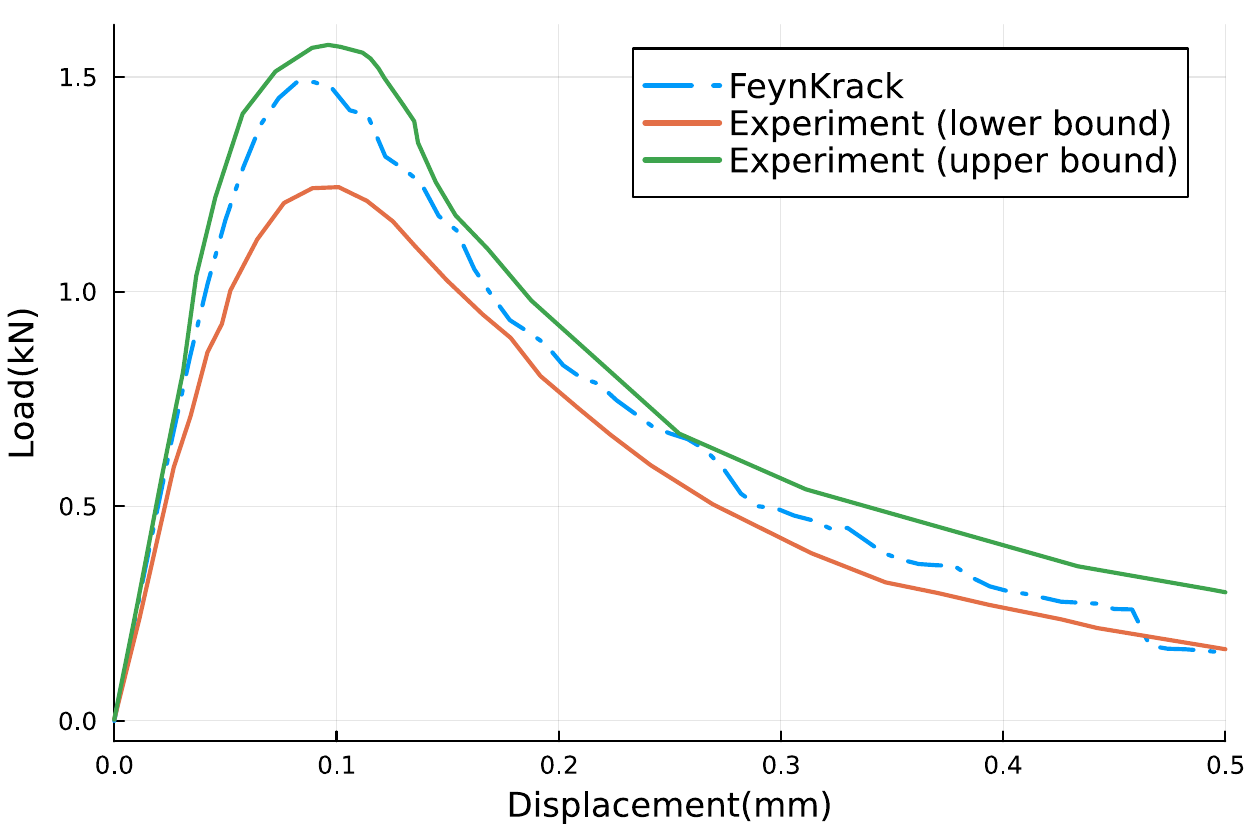}
    \caption{Comparison of load-displacement plot vis-\'a-vis  experimental data for the three-point bending test}
\label{fig:threePBendExpCompfinal}
\end{figure}

To validate FeynKrack against mode I failure, we carry out simulations on a three-point bending test on plain concrete specimens. This is a common experimental method for studying concrete failure in notched specimens. In this experiment, the crack typically originates at the notch tip and propagates vertically along the line of symmetry. The specimen's geometry along with loading and boundary conditions are depicted in Fig. \ref{fig:ThreePointBendingGeom}. The material properties for concrete vary within these ranges: Young's modulus  (18000 to 30000 MPa), Poisson's ratio (0.18-0.2), and tensile strength (2.4-3.6 MPa) \cite{winkler2001experimental, kormeling1983determination}. For the present study, we use $G_0 = 0.145 \textrm{mm}^2/N$, Young's modulus = 20000 MPa, tensile strength $\sigma_{\text{crit}}$ = 2.4 Mpa and Poisson's ratio = 0.2.
We also refine the mesh to a size of 1.5 mm near the notch and use 16 mm elsewhere (Fig. \ref{fig:ThreePBendMesh_o16_h1p5}). The domain is discretized with 2,216 elements and an effective radius of 3 mm is employed. A displacement increment of 0.008 mm is applied till a maximum displacement of 0.5 mm. The force-displacement response exhibits a characteristic linear-elastic part followed by nonlinear softening \cite{kormeling1983determination}. FeynKrack accurately predicts this behavior (see Fig. \ref{fig:threePBendExpCompfinal}). Moreover, the simulated damage distribution (crack pattern) aligns well with the experimental evidences. The contour plot of damage is shown in Fig. \ref{fig:ThreePBendCrack_Exp_comp}.

\subsection{L shaped specimen}
Mixed-mode fracture problems are more complex and predicting trajectories within a reasonable computational cost is generally challenging. We use FeynKrack to predict the crack path in an L-shaped concrete specimen subjected to a vertical concentrated load as in \cite{winkler2001experimental}.
\begin{figure}[ht!]
    \centering
\begin{subfigure}[b]{0.44\textwidth}
         \centering
\includegraphics[width=0.9\textwidth]{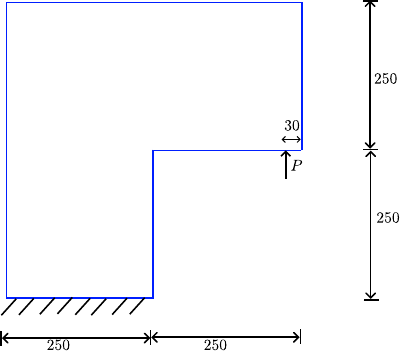}
    \caption{}
\label{fig:LShapeGeom}
\end{subfigure}
\begin{subfigure}[b]{0.44\textwidth}
         \centering
 \includegraphics[width=0.9\textwidth]{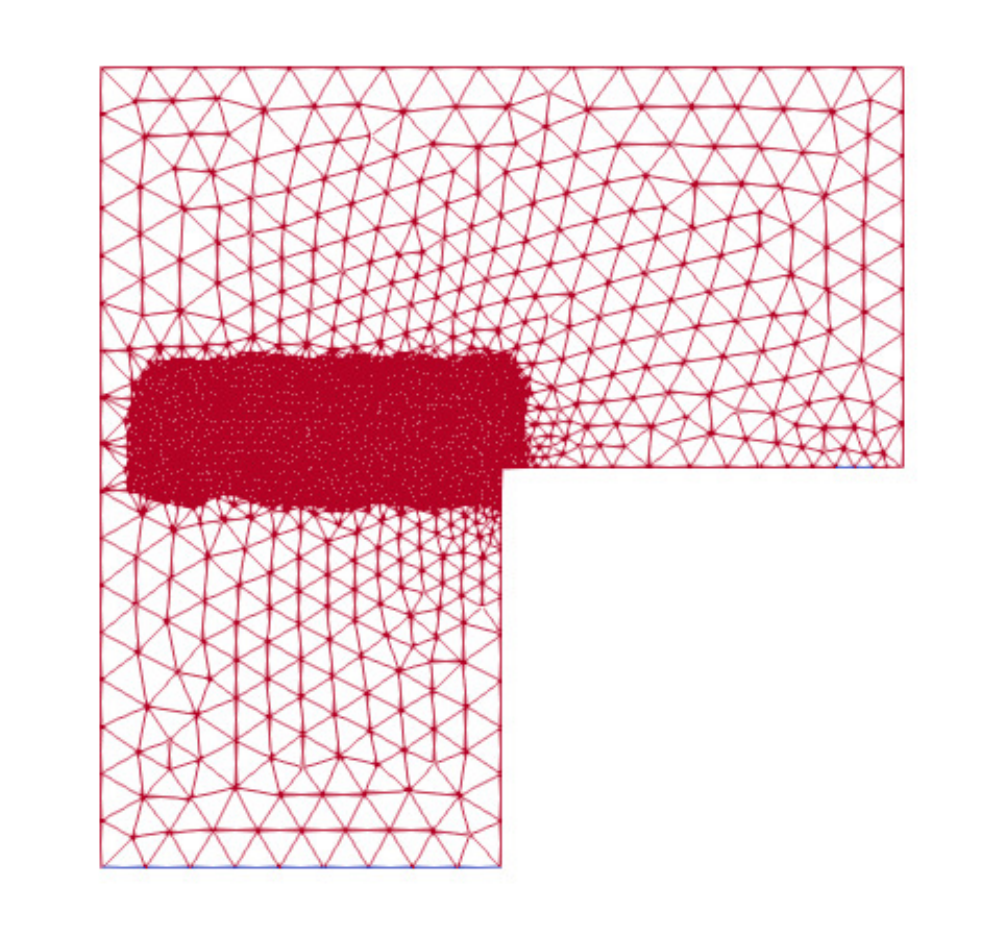}
         \caption{}
         \label{fig:Lshape_Mesh}
     \end{subfigure}
\caption{(a) Geometry and boundary conditions for L shape specimen (all dimensions are in mm); (b)  L-shaped cross section of an unnotched concrete beam: FE mesh with $m_{\text{avg}}$ = 1.5 mm around the region where the crack is expected to propagate and $m_{\text{avg}}$ = 40 mm elsewhere}
\end{figure}
Fig. \ref{fig:LShapeGeom} shows the geometry and boundary conditions of the L-shaped structural member. The following material properties are chosen: $G_0 = 0.11 \textrm{mm}^2/N$, Young's modulus $E= 22000 $ MPa, tensile strength $\sigma_{\text{crit}}$ = 2.7 MPa, and Poisson's ratio $\nu$ = 0.18. The mesh size in the vicinity of the crack is approximately 1.5 mm. It gradually increases with distance from the crack.  The specimen is subjected to a monotonically increasing displacement at the rate of 0.008 mm per increment, reaching a total displacement of 0.5 mm. The resulting damage profile is plotted in Fig. \ref{fig:Lshape_crack}, where predominant failure is observed under combined action of tension and shear. The simulated crack path closely matches with experimental observations and numerical studies (see Fig. 20 in \cite{guan2022improved}).

\begin{figure}[ht!]
    \centering
    \begin{subfigure}[b]{0.54\textwidth} 
        \centering
        \begin{minipage}[b]{0.18\textwidth}
            \raisebox{3ex}{\includegraphics[width=\textwidth]{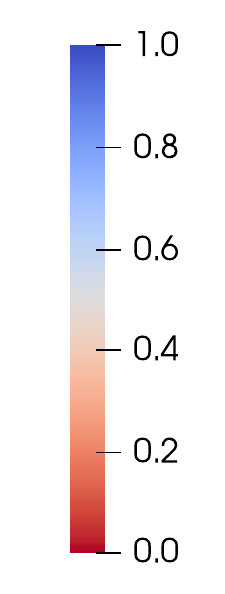}}
        \end{minipage}%
        \hfill
        \begin{minipage}[b]{0.78\textwidth}
            \centering
            \includegraphics[width=0.9\textwidth]{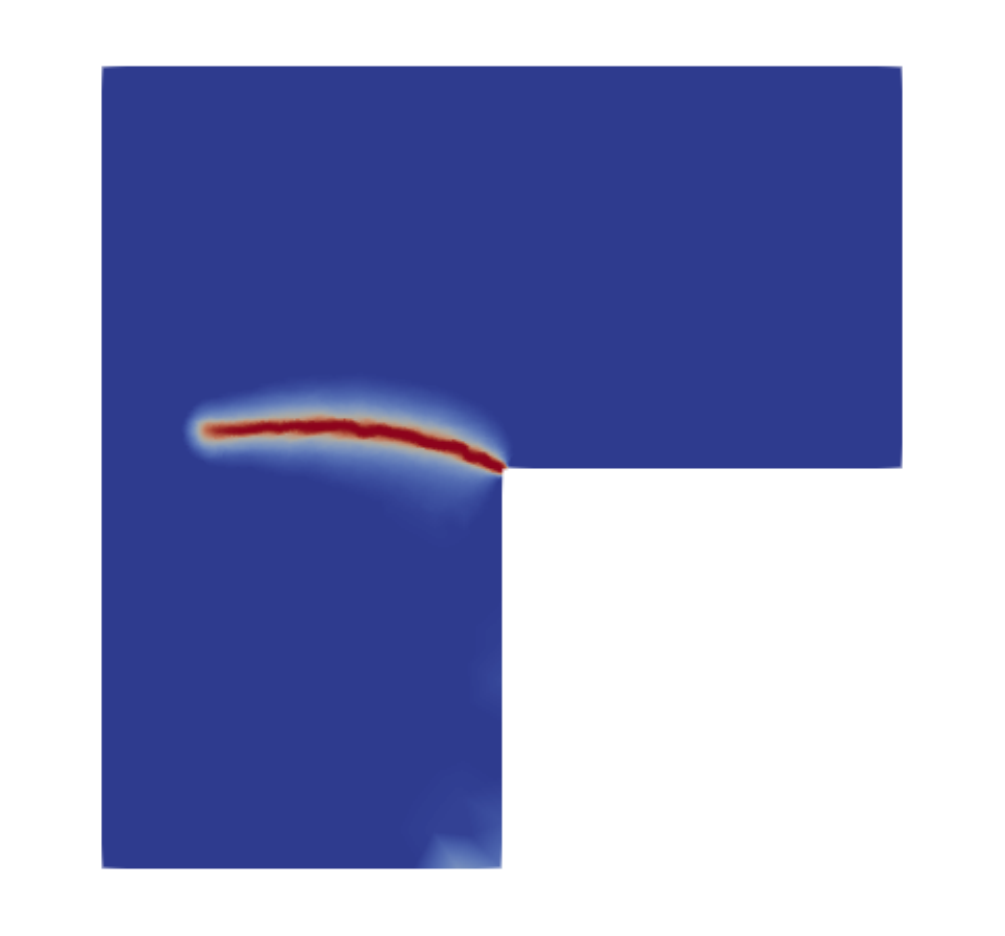}
        \end{minipage}
    \end{subfigure}
    \caption{Contour plot of damage in this study.}
    \label{fig:Lshape_crack}
\end{figure}

\begin{figure}[ht!]
    \centering
    \begin{subfigure}[b]{0.49\textwidth}
    \centering
 \includegraphics[width=0.95\textwidth]{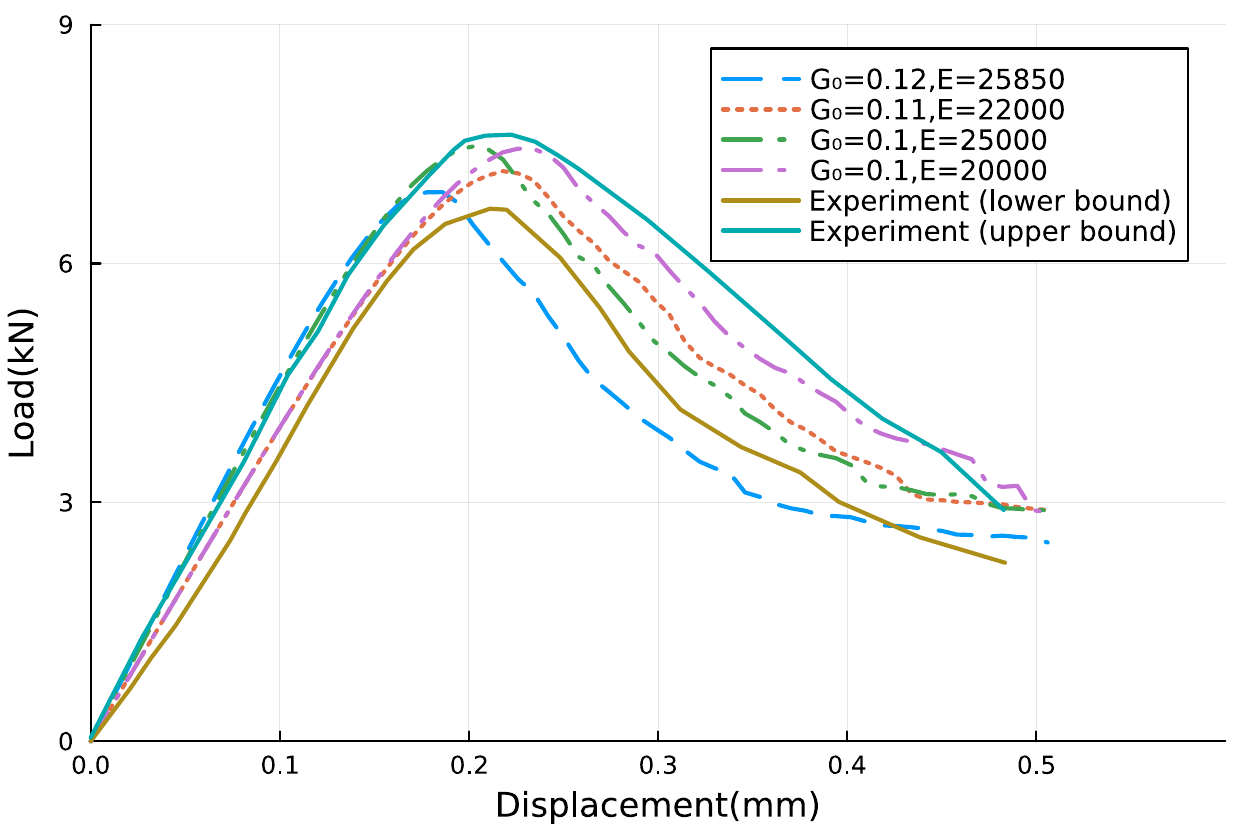}
    \caption{}
\label{fig:Lshape_expt_comp_multi}       
    \end{subfigure}
\begin{subfigure}[b]{0.49\textwidth}
\centering
\includegraphics[width=0.95\textwidth]{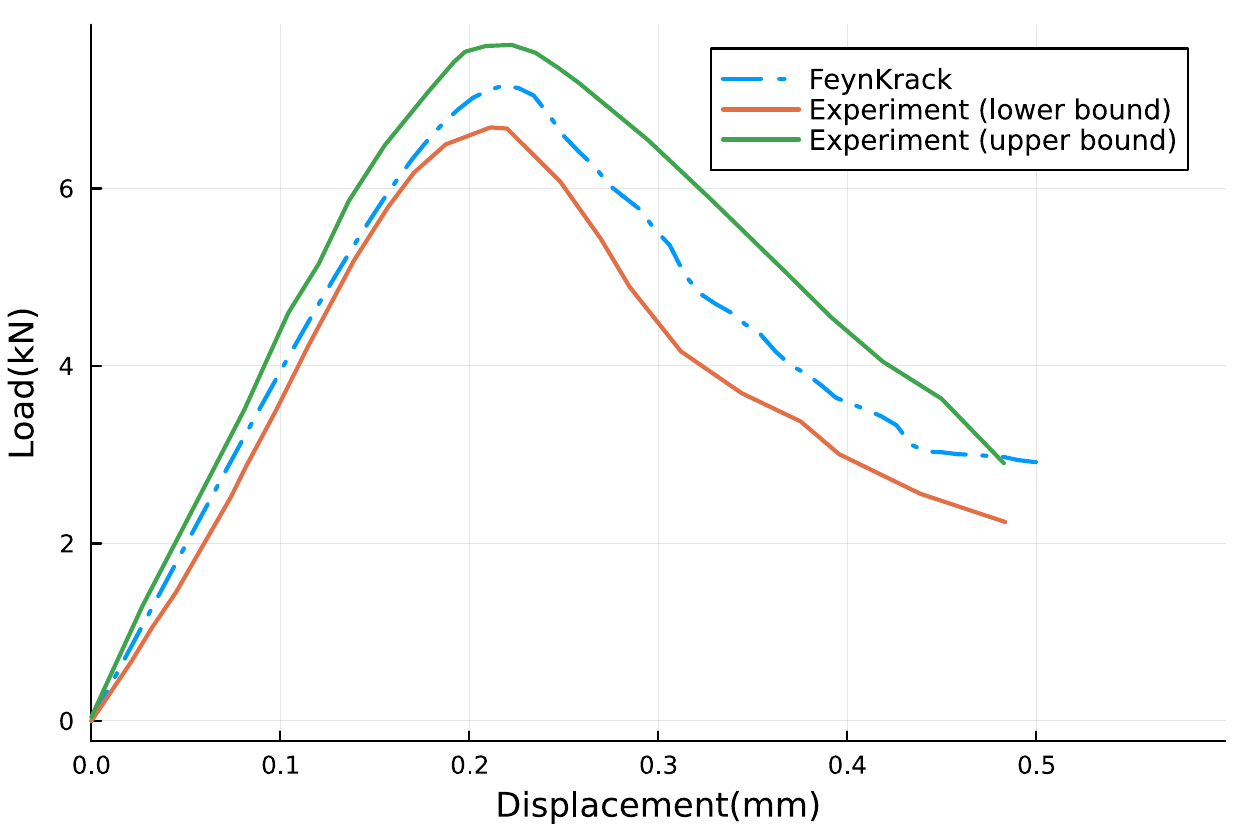}
    \caption{}
\label{fig:Lshape_expt_comp}
\end{subfigure}
\caption{(a) Comparisons of load-displacement plots via FeynKrack (with different values of parameters) with experiments on an L-shaped specimen (b) Comparisons of load-displacement plots using the FeynKrack with experiments on an L-shaped specimen; $E$ = 22000 Mpa, $\sigma_{\text{crit}}$ = 2.7 MPa, $\nu$ = 0.18}
\end{figure}

Fig. \ref{fig:Lshape_expt_comp_multi} shows force-displacement plots obtained through simulations with various values of the parameters as well as comparisons with experimental results from \cite{winkler2001experimental}. Fig. \ref{fig:Lshape_expt_comp} also shows similar results, but now with optimized values of $m_{\text{avg}}$ and $G_0$. The experimental and simulation results compare well.

\subsection{Uniaxial compression test}
\label{sec:wing cracks}
\begin{figure}[ht!]
 \centering
    \begin{subfigure}[b]{0.4\textwidth}
         \centering
\includegraphics[width=\textwidth]{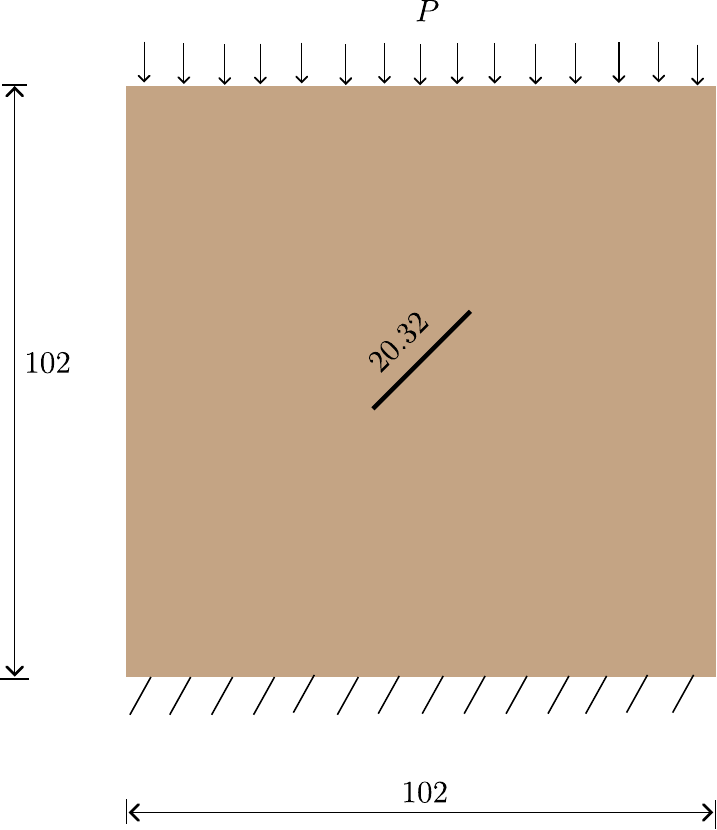}
    \caption{}
    \label{fig:UniAxcompGeomLoad}
     \end{subfigure}
    \begin{subfigure}[b]{0.4\textwidth}
         \centering
 \raisebox{4ex}{\includegraphics[width=\textwidth]{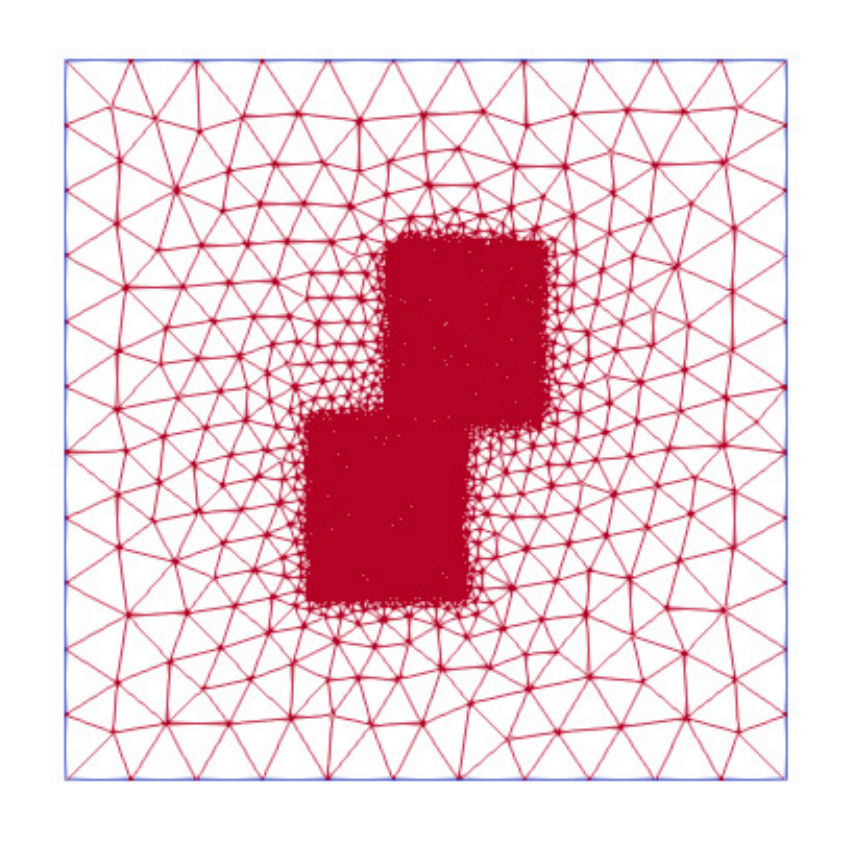}}
    \caption{}
    \label{fig:UniAxiCompMesh}
     \end{subfigure}
  \caption{(a) Geometry and boundary conditions for uniaxial compression test specimen (all dimensions are in mm), (b) mesh used for simulation (with $m_\text{avg}$ = 0.2 mm)}
  \label{fig:UniAxCompCrack2D}
\end{figure}

\begin{figure}[ht!]
    \centering
\begin{subfigure}[b]{0.4\textwidth}
         \centering
\includegraphics[width=0.8\textwidth]{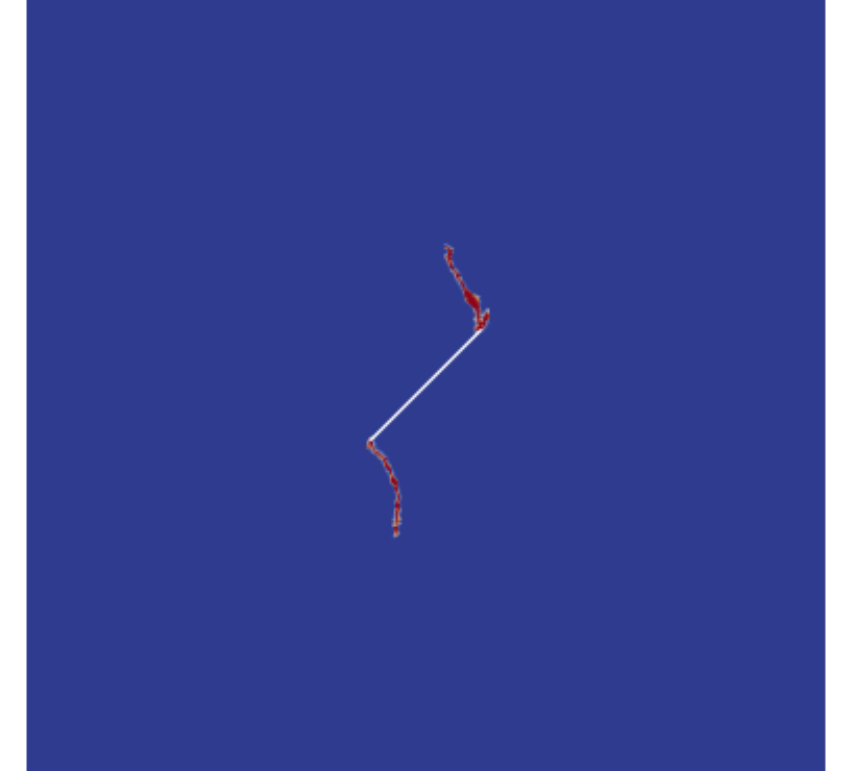}    \caption{}
\label{fig:UniAxCompCrack1}
\end{subfigure}
\begin{subfigure}[b]{0.4\textwidth}
         \centering
\includegraphics[width=0.8\textwidth]{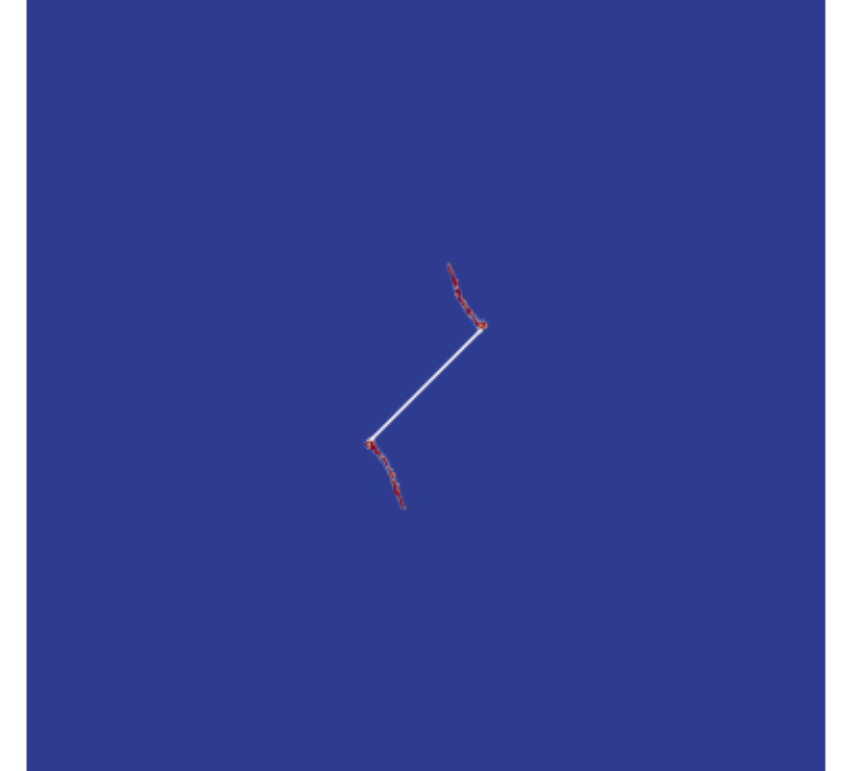}    \caption{}
\label{fig:UniAxCompCrack2}
\end{subfigure}
  \centering
    \begin{subfigure}[b]{0.1\textwidth}
         \centering
\includegraphics[width=1.6\textwidth]{colorbar_ver.pdf}
     \end{subfigure}
     \caption{\hl{Crack path in uniaxial compression; (a) $m_\text{avg} = 0.2$ mm with 63258 elements; (b) $m_\text{avg} = 0.15$ mm with 110818 elements}}
     \label{fig:UniAxCompCrack}
     \end{figure}

To assess how the model predicts \hl{mixed-mode fracture arising due to uniaxial compression and inclined crack}, we carry out simulations for a uniaxial compression test on rectangular specimens with inclined cracks. We validate our results with \cite{ingraffea1980finite}. The geometry and boundary conditions of the specimen are shown in Fig. \ref{fig:UniAxcompGeomLoad}. The inclined notch has a $45^\circ$ orientation with the horizontal. The finite element mesh is shown in Fig. \ref{fig:UniAxiCompMesh} correponds to the minimum average element size ($m_\text{avg}$) of $0.2$ mm in the refined region. The primary features of fracture here include the formation of two symmetric cracks that grow from points initially under tensile stress concentration on the notch, stable crack propagation and a curvilinear nature of propagation \cite{ingraffea1980finite}. 
We take the following material properties: $G_0 = 1.0 mm^2/N$, E = 36,200 MPa, $\sigma_{\text{crit}}$ = 10 MPa, and $\nu$ = 0.21. \hl{Refined meshes of average element sizes $0.2$ mm and $0.15$ mm are} employed over the region through which the crack is expected to propagate. The mesh outside this region uses a coarser average element size of 10 mm. Effective radius $r$ was chosen to be $0.4$ mm. The damage profile is shown in Fig. \ref{fig:UniAxCompCrack}. The model's prowess to accurately reproduce the formation of symmetric wing cracks (see Fig. 3 of \cite{ingraffea1980finite}) is demonstrated qualitatively in Fig. \ref{fig:UniAxCompCrack}. \hl{While crack paths are similar using both meshes, they are smoother with mesh refinement.}

\subsection{Three dimensional problems}

The proposed model is expected to significantly slash the computational effort whilst enabling the use of higher step-sizes and should thus be very efficacious in simulating complex three-dimensional problems. We demonstrate this through three-dimensional simulations of a wing-crack problem. We also simulate crack trajectories in the Brazilian test.

\subsubsection{Uniaxial compression test with wing-crack}
\begin{figure}[ht!]
    \centering
\includegraphics[width=0.4\textwidth]{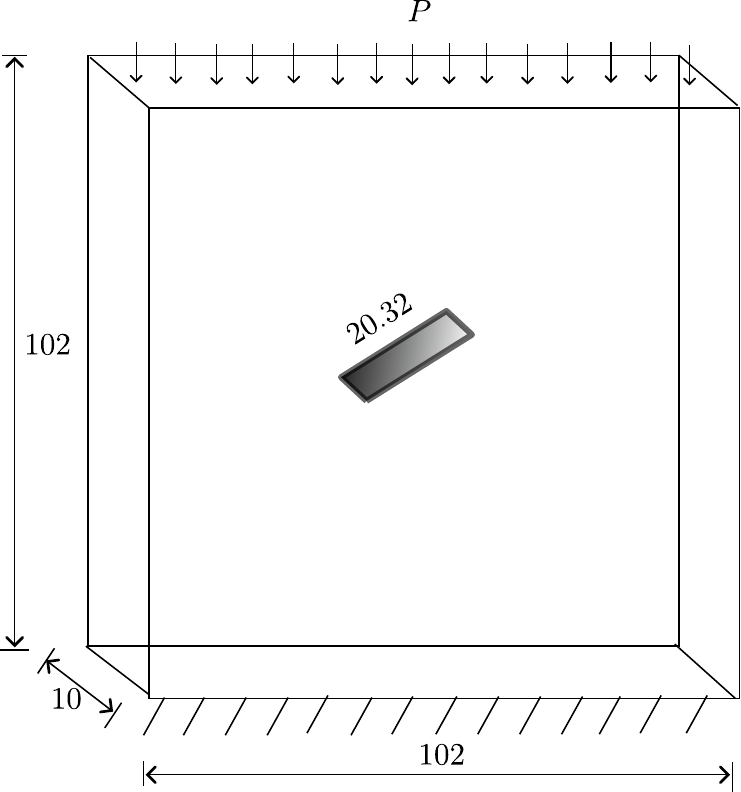}
    \caption{Geometry and boundary conditions for uniaxial compression test in 3D (all dimensions are in mm)}
\label{fig:Uniaxial_Compression_Geom3D}
\end{figure}

\begin{figure}[ht!]
 \centering
 \begin{subfigure}[b]{0.44\textwidth}
         \centering
\includegraphics[width=0.8\textwidth]{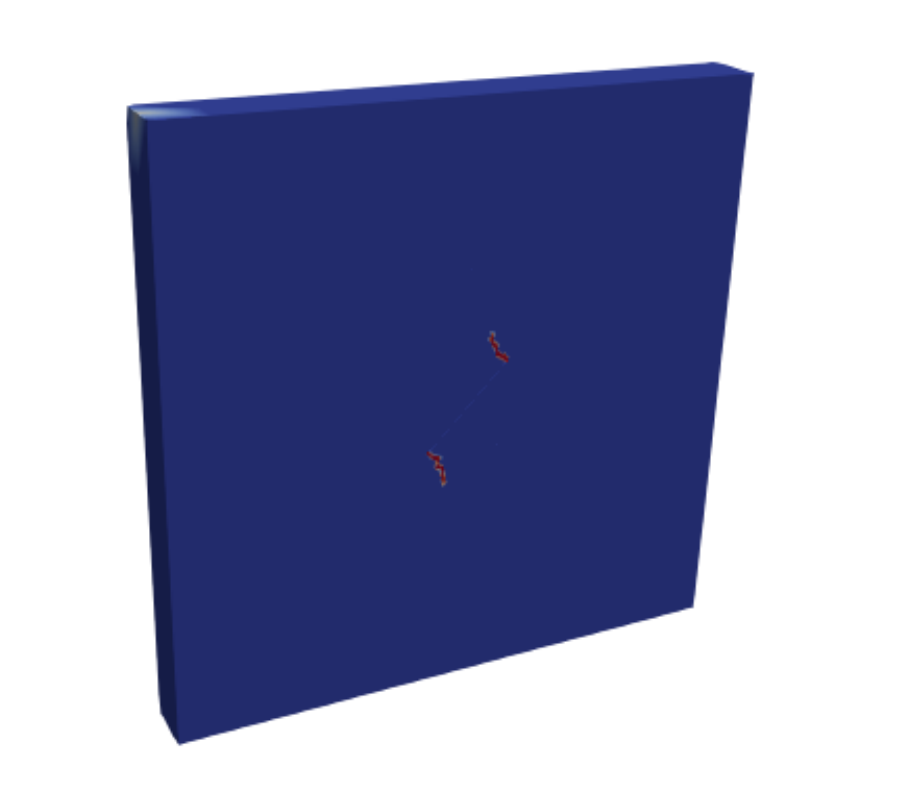}
    \caption{}
    \label{fig:UniAxiComp3DDamageIso}
     \end{subfigure}
      \begin{subfigure}[b]{0.44\textwidth}
         \centering
\includegraphics[width=0.70\textwidth]{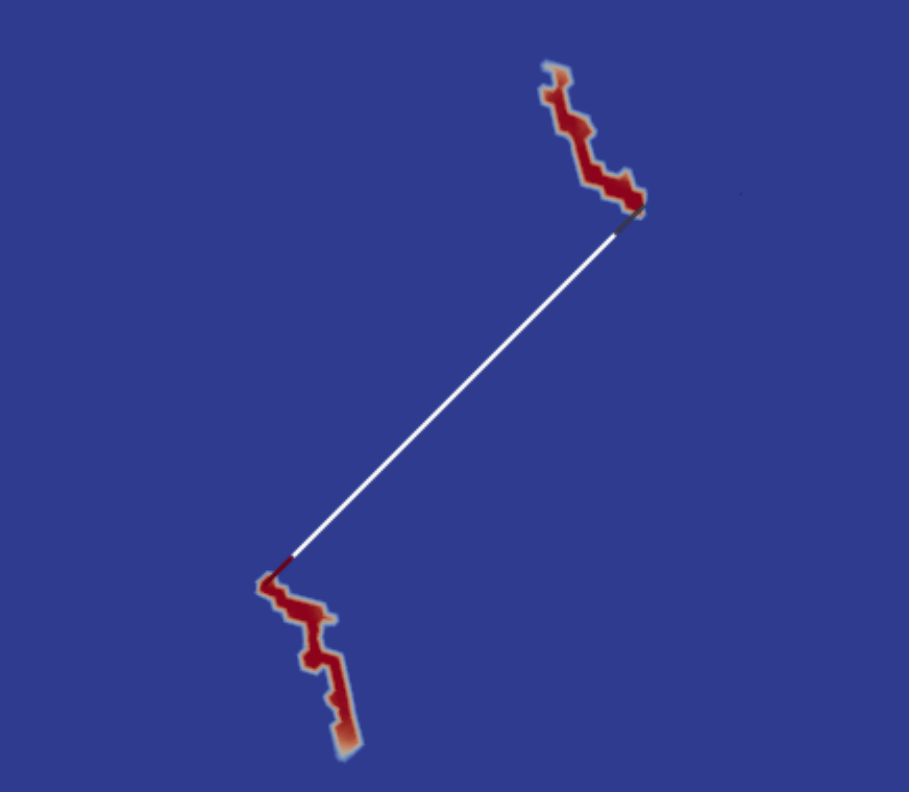}
    \caption{}
    \label{fig:UniAxiComp3DDamageIsoZoom}
     \end{subfigure}
     \centering
    \begin{subfigure}[b]{0.1\textwidth}
         \centering
\includegraphics[width=1.4\textwidth]{colorbar_ver.pdf}
     \end{subfigure}
    \begin{subfigure}[b]{0.48\textwidth}
         \centering
\includegraphics[width=0.70\textwidth]{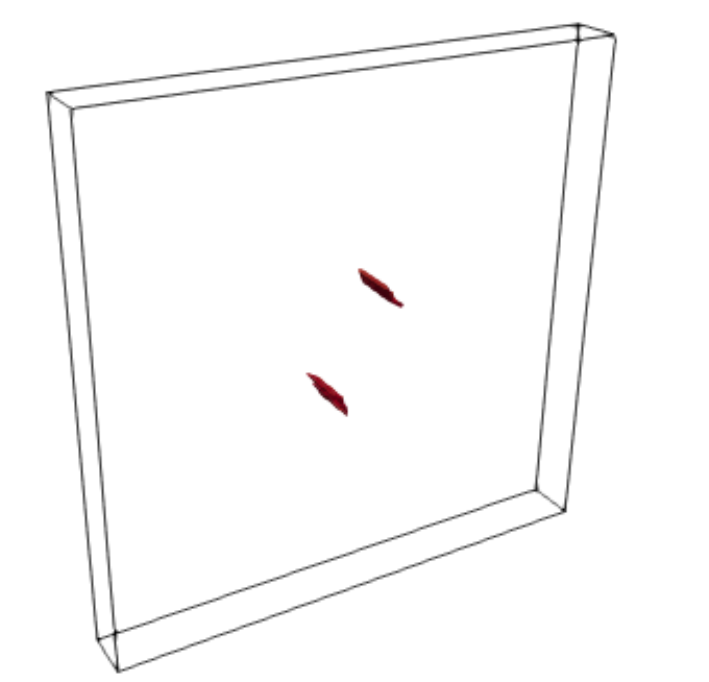}
    \caption{}
    \label{fig:UniAxiComp3DDamage}
     \end{subfigure}
    \begin{subfigure}[b]{0.48\textwidth}
         \centering
 \includegraphics[width=0.60\textwidth]{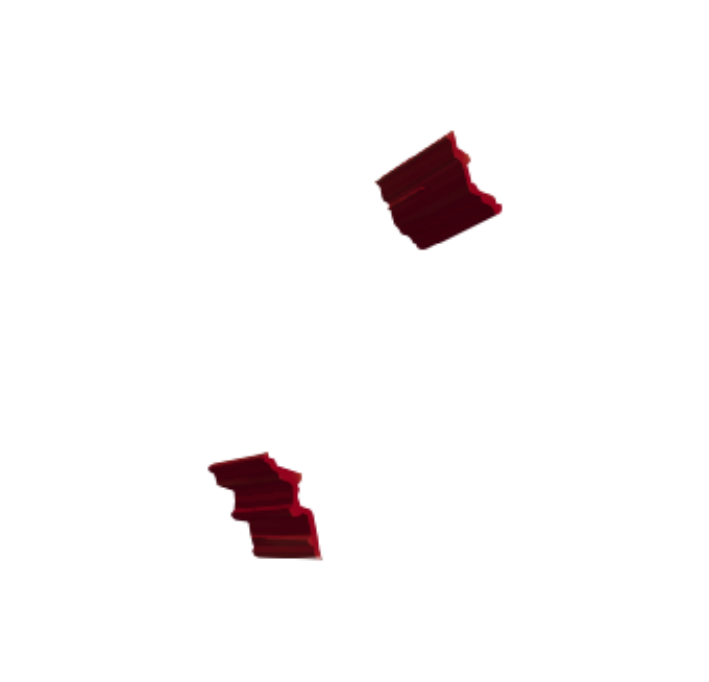}
    \caption{}
    \label{fig:UniAxiComp3DDamageZoomed}
     \end{subfigure}
  \caption{Crack path in uniaxial compression in 3D; (a) isometric view with crack surface contour; (b) zoomed view of crack (front view); (c) crack surface; (d) zoomed crack surface}
  \label{fig:UniAxCompCrack3d}
\end{figure}

 Extending the uniaxial compression test from 2D to 3D, consider a plate of constant thickness so that the inclined notch becomes a through-thickness crack, while geometry and boundary conditions remain consistent with the 2D case (\S \ref{sec:wing cracks} ). The thickness is set to 10 mm, and the same material properties are used as in \S \ref{sec:wing cracks}; see Fig. \ref{fig:Uniaxial_Compression_Geom3D}. An isometric view of the simulated crack path is in Fig. \ref{fig:UniAxiComp3DDamageIso} and a zoomed-in view in Fig. \ref{fig:UniAxiComp3DDamageIsoZoom}. The crack surface is shown in Fig. \ref{fig:UniAxiComp3DDamage} with a zoomed-in view in Fig. \ref{fig:UniAxiComp3DDamageZoomed}. The model accurately captures the crack propagation characteristics, including the formation of wing cracks.

\subsubsection{Brazilian test}
\begin{figure}[ht!]
    \centering
    \begin{subfigure}[b]{0.33\textwidth}
         \centering
\includegraphics[width=\textwidth]{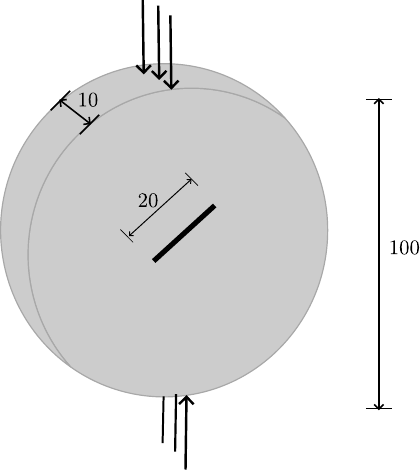}
    \caption{}
    \label{fig:BrazilianDiskGeom} 
    \end{subfigure}
    \begin{subfigure}[b]{0.33\textwidth}
        \centering
\includegraphics[width=0.9\textwidth]{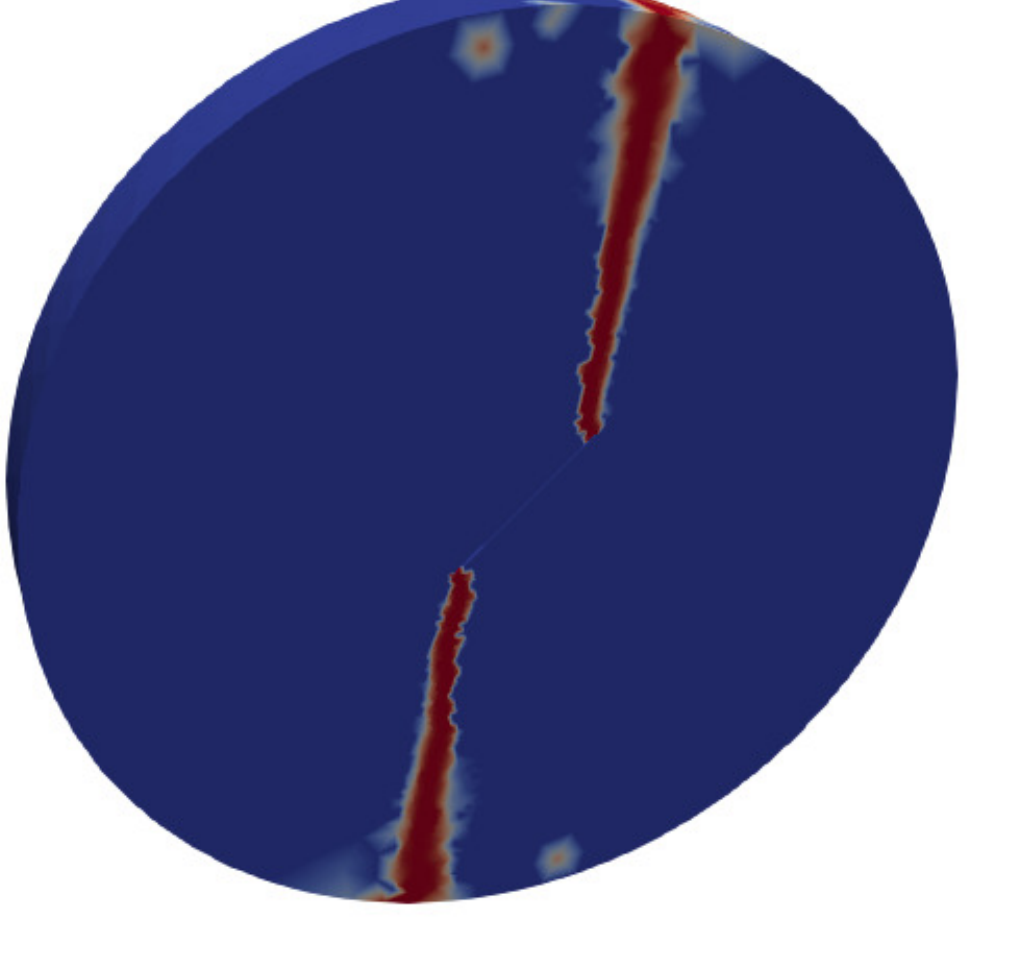}
    \caption{}
\label{fig:BrazilianDiskCrack}        
    \end{subfigure}
    \centering
    \begin{subfigure}[b]{0.1\textwidth}
         \centering
\raisebox{9ex}{\includegraphics[width=1.0\textwidth]{colorbar_ver.pdf}}
     \end{subfigure}
\caption{Brazilian disc specimen; (a) geometry and boundary conditions (all dimensions are in mm); (b) crack propagation}
\label{fig:BrazilianDisk3DGeomCrack}
\end{figure}

The Brazilian test is widely used to indirectly determine the tensile strength of rocks. It involves applying a compressive load to a cylindrical or disk-shaped specimen, causing it to fail in tension. \hl{The aim of the problem is just to show that the proposed model in its current form can capture certain experimentally observed crack propagations in 3D}. We simulate crack initiation and propagation in a Brazillian disc specimen with inclined crack. The geometry and loading condition are shown in Fig. \ref{fig:BrazilianDiskGeom}. The circular disc has a prenotch, inclined at $45^\circ$, at the center. The diameter of the disc is 100 mm and the thickness 10 mm. The material properties are:  $G_0$ = 1.0 mm$^2$/N, E = 15,000 MPa, $\sigma_{\text{crit}}$ = 3.81 Mpa, and $\nu$ = 0.21. The effective radius $r$ is taken as 1.2 mm. \hl{The minimum average element size is $0.6$ mm in the refined region and 216998 tetrahedral elements are used}. The crack path predicted by our model, as shown in Fig. \ref{fig:BrazilianDiskCrack}, aligns closely with the experimental fracture behavior (see Fig. 5(d) in \cite{haeri2014experimental}).  

\subsection{Comparison with PFM}
\label{sec:CompPFM}
\begin{figure}[ht!]
    \centering
    \begin{subfigure}[b]{0.51\textwidth}
        \centering
        \includegraphics[width=0.95\textwidth]{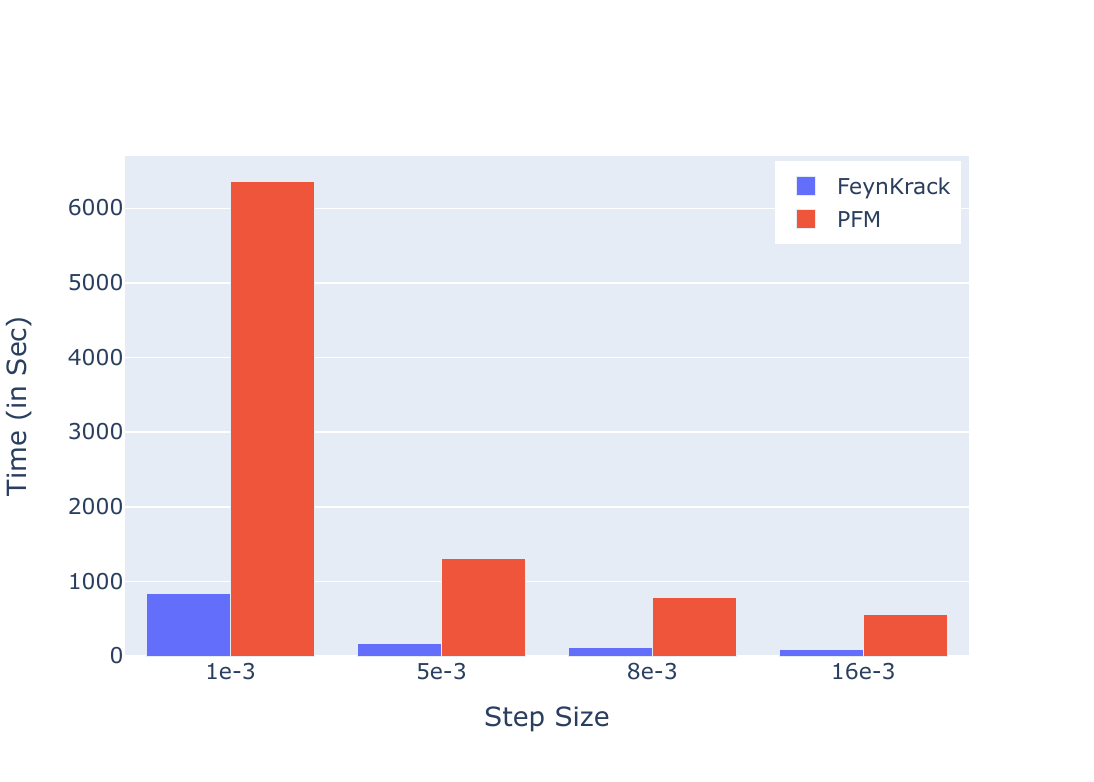}
        \caption{}
        \label{fig:Time_comparision}
    \end{subfigure}
    \hfill
    \begin{subfigure}[b]{0.45\textwidth}
        \centering
        \includegraphics[width=0.95\textwidth]{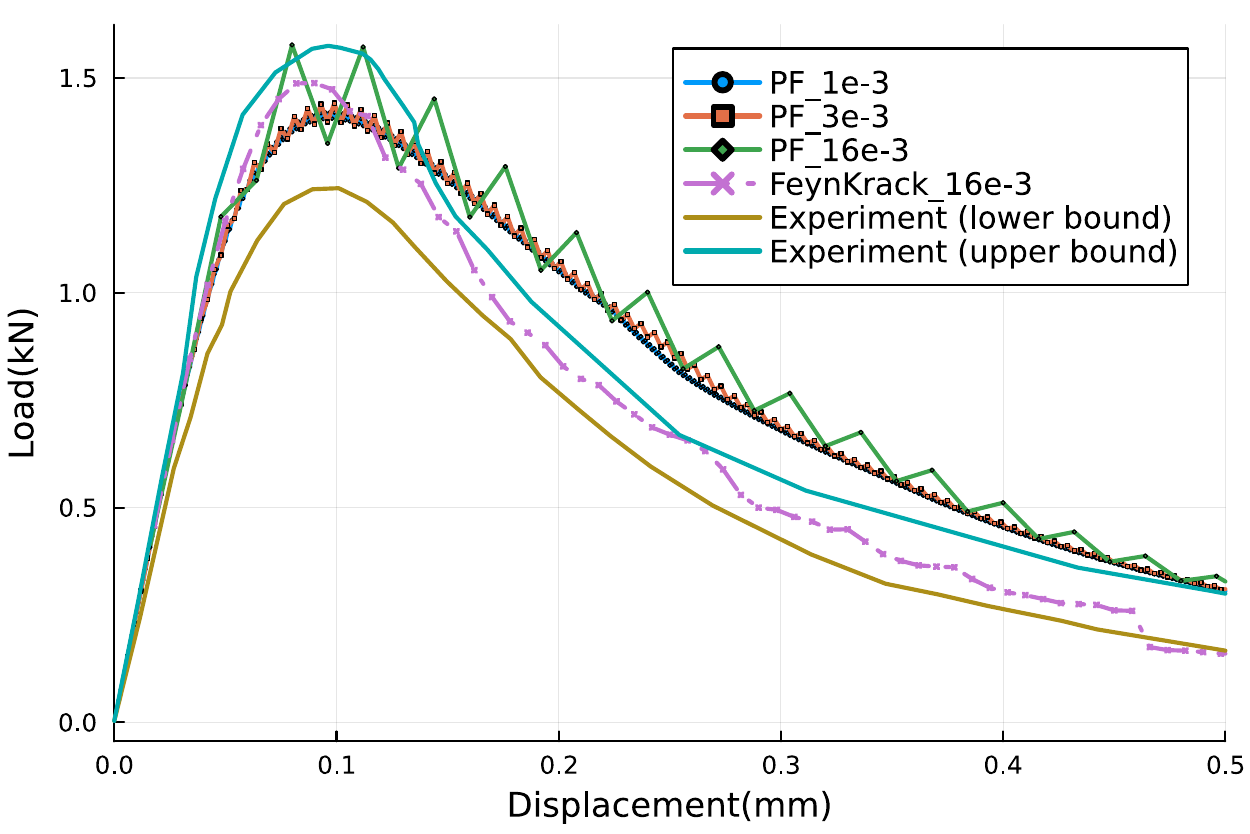}
        \caption{}
        \label{fig:Time_comparisionPdelta}
    \end{subfigure}
    \caption{(a) A comparison of CPU times required by FeynKrack and PFM; (b) Load-displacement plots for different step sizes using PFM and FeynKrack}
    \label{fig:Time_combined}
\end{figure}

We now compare the computational efficacy of the present model vis-\'a-vis the cohesive phase-field model (PFM) \cite{suh2020phase} (see \ref{sec:PFMformulation}), and the results are given in Fig. \ref{fig:Time_comparision}. Both models are implemented via Gridap \hl{and a system with Intel(R) Core(TM) i7-4930K CPU @ 3.40GHz and 16.0 GB RAM}. For the comparison, we have selected the three-point bending problem as in \S \ref{sec:ThreePBend}. For the PFM, we take $G_c = $ 0.113 N/mm, and length scale parameter $l$ = 2.5 mm. The shape parameter $p$ is assumed to be 1.0. We have compared the time taken by the two models to reach a maximum displacement of 0.5 mm. As anticipated, our results bear evidence to significant speed-up when using the present model. Specifically, the proposed model appears to be manifold faster than the specific version of PFM chosen (see Fig. \ref{fig:Time_comparision}) for identical time step-sizes. However, results via FeynKrack remain accurate for larger time steps, even as PFM's accuracy sharply deteriorates with increasing step size. Fig. \mbox{\ref{fig:Time_comparisionPdelta}} shows a comparison of load displacement plots through FeynKrack and PFM with different step-sizes. One clearly observes the robust performance of FeynKrack even with larger step-sizes. Although the maximum permissible step-size may vary depending on the problem (e.g. nonlinearity in the load-displacement curve), FeynKrack suffers no deterioration in prediction quality even with a step-size that is at least 5 times the maximum possible size for the PFM.  Assisted by the data in Fig. \ref{fig:Time_comparision}, it implies a minimum speed-up of 36-times via FeynKrack vis-\'a-vis a typical implementation of the PFM. 
However, a note of caution is warranted here. The form of \mbox{\eqref{eq:BackKolmKilDiff}} with $\boldsymbol{F} =0$  has aspects of similarity with the phase field evolution Eq. \mbox{\eqref{eq:PF_Governing_Eqn}}  (if the viscous regularization term is included in the latter), even though the former has no history dependent term and the constant source term ($3/8$ in this case). This suggests that a certain simplifying approximation, e.g., the use of a fundamental solution or Green's function, could also be introduced in the PFM to expedite computations. The heat kernel, which has the form of a zero-mean Gaussian density, could be exploited for this purpose. A similar observation has been made in \mbox{\cite{miehe2010phase}}. For a 1D case, it has been pointed out that the solution of the homogenous part of the phase field governing equation is nothing but an exponential function representing the diffusive crack topology. Beyond computational expedience, this observation also underscores the scope for a physically meaningful, measure-theoretic interpretation of the phase field, aspects that remain largely unattended to this date.

\section{Conclusions}
\label{sec:conclusion}
FeynKrack, the proposed model for quasi-brittle damage, uses a decaying measure of macroscopic bonds whose evolution is driven by the Feynman-Kac semigroup. The underlying transition density corresponding to this dynamics makes the bonds and hence the locations of material points random. This uncertainty is only to be intuitively expected when microcracks develop and propagate. A carefully crafted dynamical system for this uncertainty through a suitable choice of the killing rate is all it takes to model damage. In our stochastic setting, the dynamics is naturally non-local with the diffusion coefficient usurping the role of a length scale. As we have argued, the killed diffusion bears an asymptotic equivalence to a diffusive system with drift, which is derivable from a potential field that can be interpreted as the release rate of the fracture energy density. While the same transition bond density (for damage evolution) is used to write the macroforce (linear momentum) balance in the mean form, we have not utilized equations for higher order moments of the so-called macroforces. This has been done to align FeynKrack, as far as possible, with the CDM. From a practical standpoint, the ability to approximate the solution of the evolving transition density of bonds in closed form brings forth computational expedience over a typical implementation of the phase field method. \hl{The latter method for fracture has been developed over two decades with several speed-enhancing techniques, such as spectral solvers (e.g. FFT \mbox{\cite{chen2019fft}}). Some of these could be adapted to our method as well.}

Evaluation of the non-local quantities, in the form of expectations with respect to the transition probability, relies on the number of integration points. Zig-zag patterns observed in numerical studies may possibly be attributed to the finite number of integration points determined by domain discretization. \hl{Note that the main emphasis here is the continuous
measure-based perspective to fracture, and this naturally brings forth the so-called ’non-locality’. The covariance
of the underlying probability density parallels the support of the nonlocal kernel used in the CDM to average
the driving term. In the CDM, this finitely supported kernel acts as the localisation limiter and brings forth
mesh-insensitivity. Clearly, the use of continuous measures in our model should play a precisely similar role. Moreover, problems under compression considered here are to test the suitability of the killing term. These
are best seen as a qualitative study. To our knowledge, several useful alterations are available in CDM and phase
field models to capture fracture characteristics (e.g. of rocks) under compression. We have
managed to reproduce one of these (wing cracks) in the current setup. In any case, the form of the present killing
rate is not general enough to show the diverse features of rock fracture. Several modifications
introduced in the PFM for this purpose should provide us with the insight needed to arrive at the
correct form of killing in our future work.} Despite this, our results agree well with the experimental evidence. 

While we do not undertake it here, \hl{the effect of certain aspects of material heterogeneity
(such as coarse aggregates in concrete) on the response could be brought out by introducing a Cosserat-type
rotation, again described via a measure, within the present model}. The stochastic dynamical foundation laid out in this work may be useful in better modeling of the constitutive relations through the rich theory of non-equilibrium thermodynamics. Of specific interest in this context would be the development of an appropriate fluctuation relation to replace the second law -- the entropy inequality. Extending this idea to model ductile damage would also be a future route worth exploring. Finally, if this work has managed to go a little farther, it is only by riding 'on the shoulders of giants', such as the originators of schemes like phase field \cite{miehe2010phase} and GraFEA \cite{thamburaja2021fracture}.

\appendix
\section{Phase field model for cohesive fracture}
\label{sec:PFMformulation}
This section summarises the governing equations for PFM \cite{suh2020phase} used in the \S \ref{sec:CompPFM}.
The linear momentum balance equation is as follows.
\begin{subequations}
\begin{align}
     \nabla \cdot w(s)\boldsymbol{\sigma} = \boldsymbol{0}\,\, \text{on}\,\, \Lambda_0 \\
  \boldsymbol{\sigma}\boldsymbol{n} = \boldsymbol{t}  \,\, \text{on}\,\, \Gamma_t   
\end{align}
\end{subequations}
Here the degradation function $w(s)$ is assumed to be of the form:
\begin{equation}
    w(s) = \frac{s^2}{s^2+m(1-s)(1+p(1-s))}\,\, \text{with}\,\, l \le \frac{3 G_c}{8(p+2)\psi_{crit}} \,\, \text{and}\,\, m = \frac{3}{8 \mathcal{F}_{crit}}
    \label{eq:PFDegFunc}
\end{equation}
$p$ and $m$ are the parameters controlling the degradation rate during the damage process. $G_c$ is the critical energy release rate. 

The governing equation corresponding to damage evolution is given by:
\begin{equation}
        \nabla_{\boldsymbol{X}} \cdot \left(\frac{3}{4}l^2\nabla_{\boldsymbol{X}} s\right)-w'(s)\mathcal{H}+\frac{3}{8} = M \dot{s}
     \label{eq:PF_Governing_Eqn}
\end{equation}
where $\mathcal{F} = \frac{\psi^\text{d}}{G_c/l}$. $M = 0$ has been assumed for comparison in \S \mbox{\ref{sec:CompPFM}}
\begin{equation}\label{eq:PFEvolution}
    \mathcal{H} = \max_{\tau\in[0,t]}\left(\mathcal{F}_{\text{crit}}+\langle \frac{\mathcal{F}}{\mathcal{F}_{\text{crit}}}-1\rangle_+\right)
\end{equation}
The history function above ensures an initial elastic response. Evolution of damage starts only when $\mathcal{F}$ crosses $\mathcal{F}_\text{crit}$. Note that, while implementing the PFM, we have used a hybrid formulation.

\bibliographystyle{elsarticle-num-names}
\bibliography{References.bib}

\end{document}